\documentclass[reqno,12pt,a4paper]{amsart}

\usepackage{mathrsfs}
\usepackage{amssymb}
\usepackage{amsfonts}
\usepackage{latexsym}
\usepackage{amsthm}

\usepackage{graphicx}
\def\lb{\label}

\newcommand{\er}[1]{\textrm{(\ref{#1})}}

\begin{document}


\renewcommand{\theequation}{\arabic{section}.\arabic{equation}}
\theoremstyle{plain}
\newtheorem{theorem}{\bf Theorem}[section]
\newtheorem{lemma}[theorem]{\bf Lemma}
\newtheorem{corollary}[theorem]{\bf Corollary}
\newtheorem{proposition}[theorem]{\bf Proposition}
\newtheorem{definition}[theorem]{\bf Definition}
\newtheorem{remark}[theorem]{\it Remark}

\def\a{\alpha}  \def\cA{{\mathcal A}}     \def\bA{{\bf A}}  \def\mA{{\mathscr A}}
\def\b{\beta}   \def\cB{{\mathcal B}}     \def\bB{{\bf B}}  \def\mB{{\mathscr B}}
\def\g{\gamma}  \def\cC{{\mathcal C}}     \def\bC{{\bf C}}  \def\mC{{\mathscr C}}
\def\G{\Gamma}  \def\cD{{\mathcal D}}     \def\bD{{\bf D}}  \def\mD{{\mathscr D}}
\def\d{\delta}  \def\cE{{\mathcal E}}     \def\bE{{\bf E}}  \def\mE{{\mathscr E}}
\def\D{\Delta}  \def\cF{{\mathcal F}}     \def\bF{{\bf F}}  \def\mF{{\mathscr F}}
\def\c{\chi}    \def\cG{{\mathcal G}}     \def\bG{{\bf G}}  \def\mG{{\mathscr G}}
\def\z{\zeta}   \def\cH{{\mathcal H}}     \def\bH{{\bf H}}  \def\mH{{\mathscr H}}
\def\e{\eta}    \def\cI{{\mathcal I}}     \def\bI{{\bf I}}  \def\mI{{\mathscr I}}
\def\p{\psi}    \def\cJ{{\mathcal J}}     \def\bJ{{\bf J}}  \def\mJ{{\mathscr J}}
\def\vT{\Theta} \def\cK{{\mathcal K}}     \def\bK{{\bf K}}  \def\mK{{\mathscr K}}
\def\k{\kappa}  \def\cL{{\mathcal L}}     \def\bL{{\bf L}}  \def\mL{{\mathscr L}}
\def\l{\lambda} \def\cM{{\mathcal M}}     \def\bM{{\bf M}}  \def\mM{{\mathscr M}}
\def\L{\Lambda} \def\cN{{\mathcal N}}     \def\bN{{\bf N}}  \def\mN{{\mathscr N}}
\def\m{\mu}     \def\cO{{\mathcal O}}     \def\bO{{\bf O}}  \def\mO{{\mathscr O}}
\def\n{\nu}     \def\cP{{\mathcal P}}     \def\bP{{\bf P}}  \def\mP{{\mathscr P}}
\def\r{\rho}    \def\cQ{{\mathcal Q}}     \def\bQ{{\bf Q}}  \def\mQ{{\mathscr Q}}
\def\s{\sigma}  \def\cR{{\mathcal R}}     \def\bR{{\bf R}}  \def\mR{{\mathscr R}}
                \def\cS{{\mathcal S}}     \def\bS{{\bf S}}  \def\mS{{\mathscr S}}
\def\t{\tau}    \def\cT{{\mathcal T}}     \def\bT{{\bf T}}  \def\mT{{\mathscr T}}
\def\f{\phi}    \def\cU{{\mathcal U}}     \def\bU{{\bf U}}  \def\mU{{\mathscr U}}
\def\F{\Phi}    \def\cV{{\mathcal V}}     \def\bV{{\bf V}}  \def\mV{{\mathscr V}}
\def\P{\Psi}    \def\cW{{\mathcal W}}     \def\bW{{\bf W}}  \def\mW{{\mathscr W}}
\def\o{\omega}  \def\cX{{\mathcal X}}     \def\bX{{\bf X}}  \def\mX{{\mathscr X}}
\def\x{\xi}     \def\cY{{\mathcal Y}}     \def\bY{{\bf Y}}  \def\mY{{\mathscr Y}}
\def\X{\Xi}     \def\cZ{{\mathcal Z}}     \def\bZ{{\bf Z}}  \def\mZ{{\mathscr Z}}
\def\O{\Omega}

\newcommand{\gA}{\mathfrak{A}}          \newcommand{\ga}{\mathfrak{a}}
\newcommand{\gB}{\mathfrak{B}}          \newcommand{\gb}{\mathfrak{b}}
\newcommand{\gC}{\mathfrak{C}}          \newcommand{\gc}{\mathfrak{c}}
\newcommand{\gD}{\mathfrak{D}}          \newcommand{\gd}{\mathfrak{d}}
\newcommand{\gE}{\mathfrak{E}}
\newcommand{\gF}{\mathfrak{F}}           \newcommand{\gf}{\mathfrak{f}}
\newcommand{\gG}{\mathfrak{G}}           
\newcommand{\gH}{\mathfrak{H}}           \newcommand{\gh}{\mathfrak{h}}
\newcommand{\gI}{\mathfrak{I}}           \newcommand{\gi}{\mathfrak{i}}
\newcommand{\gJ}{\mathfrak{J}}           \newcommand{\gj}{\mathfrak{j}}
\newcommand{\gK}{\mathfrak{K}}            \newcommand{\gk}{\mathfrak{k}}
\newcommand{\gL}{\mathfrak{L}}            \newcommand{\gl}{\mathfrak{l}}
\newcommand{\gM}{\mathfrak{M}}            \newcommand{\gm}{\mathfrak{m}}
\newcommand{\gN}{\mathfrak{N}}            \newcommand{\gn}{\mathfrak{n}}
\newcommand{\gO}{\mathfrak{O}}
\newcommand{\gP}{\mathfrak{P}}             \newcommand{\gp}{\mathfrak{p}}
\newcommand{\gQ}{\mathfrak{Q}}             \newcommand{\gq}{\mathfrak{q}}
\newcommand{\gR}{\mathfrak{R}}             \newcommand{\gr}{\mathfrak{r}}
\newcommand{\gS}{\mathfrak{S}}              \newcommand{\gs}{\mathfrak{s}}
\newcommand{\gT}{\mathfrak{T}}             \newcommand{\gt}{\mathfrak{t}}
\newcommand{\gU}{\mathfrak{U}}             \newcommand{\gu}{\mathfrak{u}}
\newcommand{\gV}{\mathfrak{V}}             \newcommand{\gv}{\mathfrak{v}}
\newcommand{\gW}{\mathfrak{W}}             \newcommand{\gw}{\mathfrak{w}}
\newcommand{\gX}{\mathfrak{X}}               \newcommand{\gx}{\mathfrak{x}}
\newcommand{\gY}{\mathfrak{Y}}              \newcommand{\gy}{\mathfrak{y}}
\newcommand{\gZ}{\mathfrak{Z}}             \newcommand{\gz}{\mathfrak{z}}

\def\ve{\varepsilon} \def\vt{\vartheta} \def\vp{\varphi}  \def\vk{\varkappa}

\def\Z{{\mathbb Z}} \def\R{{\mathbb R}} \def\C{{\mathbb C}}  \def\K{{\mathbb K}}
\def\T{{\mathbb T}} \def\N{{\mathbb N}} \def\dD{{\mathbb D}} \def\S{{\mathbb S}}
\def\B{{\mathbb B}}


\def\la{\leftarrow}              \def\ra{\rightarrow}     \def\Ra{\Rightarrow}
\def\ua{\uparrow}                \def\da{\downarrow}
\def\lra{\leftrightarrow}        \def\Lra{\Leftrightarrow}
\newcommand{\abs}[1]{\lvert#1\rvert}
\newcommand{\br}[1]{\left(#1\right)}

\def\lan{\langle} \def\ran{\rangle}


\def\lt{\biggl}                  \def\rt{\biggr}
\def\ol{\overline}               \def\wt{\widetilde}
\def\no{\noindent}


\let\ge\geqslant                 \let\le\leqslant
\def\lan{\langle}                \def\ran{\rangle}
\def\/{\over}                    \def\iy{\infty}
\def\sm{\setminus}               \def\es{\emptyset}
\def\ss{\subset}                 \def\ts{\times}
\def\pa{\partial}                \def\os{\oplus}
\def\om{\ominus}                 \def\ev{\equiv}
\def\iint{\int\!\!\!\int}        \def\iintt{\mathop{\int\!\!\int\!\!\dots\!\!\int}\limits}
\def\el2{\ell^{\,2}}             \def\1{1\!\!1}
\def\sh{\sharp}
\def\wh{\widehat}
\def\bs{\backslash}
\def\na{\nabla}

\def\ch{\mathop{\mathrm{ch}}\nolimits}
\def\sh{\mathop{\mathrm{sh}}\nolimits}
\def\all{\mathop{\mathrm{all}}\nolimits}
\def\Area{\mathop{\mathrm{Area}}\nolimits}
\def\arg{\mathop{\mathrm{arg}}\nolimits}
\def\const{\mathop{\mathrm{const}}\nolimits}
\def\det{\mathop{\mathrm{det}}\nolimits}
\def\diag{\mathop{\mathrm{diag}}\nolimits}
\def\diam{\mathop{\mathrm{diam}}\nolimits}
\def\dim{\mathop{\mathrm{dim}}\nolimits}
\def\dist{\mathop{\mathrm{dist}}\nolimits}
\def\Im{\mathop{\mathrm{Im}}\nolimits}
\def\Iso{\mathop{\mathrm{Iso}}\nolimits}
\def\Ker{\mathop{\mathrm{Ker}}\nolimits}
\def\Lip{\mathop{\mathrm{Lip}}\nolimits}
\def\rank{\mathop{\mathrm{rank}}\limits}
\def\Ran{\mathop{\mathrm{Ran}}\nolimits}
\def\Re{\mathop{\mathrm{Re}}\nolimits}
\def\Res{\mathop{\mathrm{Res}}\nolimits}
\def\res{\mathop{\mathrm{res}}\limits}
\def\sign{\mathop{\mathrm{sign}}\nolimits}
\def\span{\mathop{\mathrm{span}}\nolimits}
\def\supp{\mathop{\mathrm{supp}}\nolimits}
\def\Tr{\mathop{\mathrm{Tr}}\nolimits}
\def\BBox{\hspace{1mm}\vrule height6pt width5.5pt depth0pt \hspace{6pt}}
\def\where{\mathop{\mathrm{where}}\nolimits}
\def\as{\mathop{\mathrm{as}}\nolimits}
\def\Dom{\mathop{\mathrm{Dom}}\nolimits}


\newcommand\nh[2]{\widehat{#1}\vphantom{#1}^{(#2)}}
\def\dia{\diamond}

\def\Oplus{\bigoplus\nolimits}



\def\qqq{\qquad}
\def\qq{\quad}
\let\ge\geqslant
\let\le\leqslant
\let\geq\geqslant
\let\leq\leqslant
\newcommand{\ca}{\begin{cases}}
\newcommand{\ac}{\end{cases}}
\newcommand{\ma}{\begin{pmatrix}}
\newcommand{\am}{\end{pmatrix}}
\renewcommand{\[}{\begin{equation}}
\renewcommand{\]}{\end{equation}}
\def\eq{\begin{equation}}
\def\qe{\end{equation}}
\def\[{\begin{equation}}
\def\bu{\bullet}

\newcommand{\fr}{\frac}
\newcommand{\tf}{\tfrac}

\title[{Resonances for Euler-Bernoulli operator}]
{Resonances for Euler-Bernoulli operator on the half-line}

\date{\today}
\author[Andrey Badanin]{Andrey Badanin}
\author[Evgeny Korotyaev]{Evgeny L. Korotyaev}
\address{Saint-Petersburg
State University, Universitetskaya nab. 7/9, St. Petersburg, 199034
Russia, an.badanin@gmail.com, a.badanin@spbu.ru,
korotyaev@gmail.com,  e.korotyaev@spbu.ru}


\begin{abstract}
\no We consider resonances for fourth order differential operators
on the half-line with compactly supported  coefficients. We
determine asymptotics of a counting function of resonances in
complex discs at large radius, describe the forbidden domain for
resonances  and obtain trace formulas in terms of resonances. We
apply these results to the Euler-Bernoulli operator on the
half-line. The coefficients of this operator are positive and
constants outside a finite interval. We show that this operator does
not have any eigenvalues and resonances iff its coefficients are
constants on the whole half-line.

\end{abstract}

\maketitle

\begin{quotation}

\begin{center}
{\bf Table of Contents}
\end{center}

\vskip 6pt

{\footnotesize

1. Introduction and main results \hfill \pageref{Sec1}\ \ \ \ \

2. Properties of the free resolvent   \hfill \pageref{Sec2}\ \ \ \ \

3. The scattering matrix and the Fredholm determinant \hfill
\pageref{Sec3}\ \ \ \ \

4. Proof of the main Theorems and trace formulas in terms of
resonances \hfill \pageref{Sec4}\ \ \ \ \

5. Euler-Bernoulli operators and proof of Theorem~\ref{ThEB}
 \hfill \pageref{Sec5}\ \ \ \ \

6. Resonances for coefficients with jump
discontinuity and proof of  Theorem~\ref{ThAsCF} \hfill \pageref{Sec6}\ \ \ \ \

}
\end{quotation}


\vskip 0.25cm

\section {\lb{Sec1} Introduction and main results}
\setcounter{equation}{0}

\subsection{Introduction}

There are many results about Schr\"odinger operators ${\bf
H}=-\D+V({\bf x}), {\bf x}\in \R^3$ with compactly supported potentials
$V$ on $L^2(\R^3)$, see \cite{Z89}, \cite{SZ91} and references
therein. In the important physical case the potential $V(|{\bf x}|)$
is symmetric and depends only on radius $r=|{\bf x}|>0$.
The standard transform $y({\bf x})\mapsto ry({\bf x})$
and expansion in spherical harmonics give that ${\bf H}$ is
unitarily equivalent to a direct sum of the Schr\"odinger operators
acting on $L^2(\R_+)$. The first operator from this sum is given by
$-{\pa^2\/\pa r^2}+V(r)$. There are a lot of results about the
resonances for 1-dimensional case, see \cite{F97}, \cite{K04},
\cite{S00}, \cite{Z87} and references therein.

Now we consider a biharmonic type operator ${\bf
B}={1\/\b(r)}\D\a(r)\D$ on $\R^3$, where $\a(r),\b(r)$ are some
positive functions depending on the radius $r$ only. The similar operators
on $\R^2$ describe, for example, vibrations of plates with an
axisymmetric variable thickness, see \cite{L69}. The separation of
variables (similar to the case of Schr\"odinger operators), show
that the operator ${\bf B}$  is unitarily equivalent to a direct sum
of fourth order operators acting on $L^2(\R_+)$. The first operator
from this sum is given by an Euler-Bernoulli operator
${1\/b}{d^2\/dr^2}a {d^2\/dr^2}$ on the half-line with some positive
coefficients $a,b$. Remark that  the Euler-Bernoulli operators are
related with the problems of vibrations of beams, see \cite{TW59}.

The standard unitary Liouville type transformation reduces the
Euler-Bernoulli operator into a fourth order operator $H$ on the
half-line defined by \er{a.1} with some coefficients $p,q$. Thus in
order to discuss resonances for Euler-Bernoulli operators we
consider a fourth order operators $H$ acting on $L^2(\R_+)$ and
given by
\[
\label{a.1}
Hy=H_0y+Vy, \qq H_0=\pa^4,\qqq V=2\pa p\pa +q,
\]
 with the boundary conditions
\[
\lb{4g.dc}
y(0)=y''(0)=0,
\]
where $\pa ={d\/dx}$. We consider the operator $H_0=\pa^4$ as unperturbed
and the operator  $V$ is its perturbation.
The coefficients $p,q$ are compactly supported
and belong to the space $\cH_0$, where
$\cH_m=\cH_m(\g),m=0,1,2,...$, is the spaces of functions defined by
$$
\cH_m=\big\{f\in L_{real}^1(\R_+): \supp f\in[0,\g],f^{(m)}\in
L^1(0,\g)\big\}
$$
for some $\g>0$.  The  boundary conditions
\er{4g.dc} are taken for reasons of convenience. The operators with
other boundary conditions can be considered similarly.

It is well known that the operator $H$ is self-adjoint and is
defined on the corresponding form domain, see Sect.~\ref{ssY0Y}.
It has purely absolutely continuous spectrum $[0,\iy)$
plus a finite number of simple real eigenvalues, see Proposition
~\ref{T1}.

\subsection{Schr\"odinger operators}
Before discussion about resonances for the fourth order operators we
recall the well known results for Schr\"odinger operators with
 compactly supported potentials on the half-line.
 For $p\in \cH_0$
we define a Schr\"odinger operator $h$ on $L^2(\R_+)$ by
\[
\lb{Schr}  h=h_0-p, \qqq\where\qq h_0f=-f'',\qq f(0)=0.
\]
Here the operator $h_0$ is  unperturbed.
The operator $h$ has purely absolutely continuous spectrum $[0,\iy)$
plus a finite number of simple negative eigenvalues
$e_1<e_2<...<e_N<0$. Define an operator
$$
y_0(k)=p^{1\/2}(h_0-k^2)^{-1}|p|^{1\/2},\qqq
p^{1\/2}=|p|^{1\/2}\sign p,\qqq k\in\C_+.
$$
Each operator $y_0(k)$ is trace class and is analytic in $k\in \C$.
Thus we can define the Fredholm determinant
\[
\lb{dt2} d(k)=\det\big(1-y_0(k)\big),\qqq k\in\C.
\]
It is well known, see, e.g. \cite{F63}, that the function $d(k)$ has
a finite number of simple zeros $\sqrt e_1,..., \sqrt e_N$ in
$\C_+$, maybe simple zero at $k=0$ and an infinite number of zeros
(resonances) in $\C_-$, see Fig.~\ref{fighsq}~a). We define
resonances as zeros of a Fredholm determinant $d(k)$ in $\C_-\cup
\{0\}$.

There are a lot of different results about resonances for
1-dimensional Schr\"odinger operators with compactly supported
potentials, see Froese \cite{F97}, Hitrik \cite{H99}, Korotyaev
\cite{K04}, Simon \cite{S00}, Zworski \cite{Z87} and references
therein. Recall the following results:

1) The set of resonances is symmetric with respect to the imaginary
axis, since the operator $h$ is self-adjoint.

2) The resonances may have any multiplicity (see \cite{K04}).

3) Let  $\g=\sup (\supp(p))$ and let  $n(r)$ be the number  of zeros
of $d(k)$ in a disk $|k|<r$. Zworski \cite{Z87} determined the
following asymptotics (see also \cite{F97}, \cite{K04}, \cite{S00})
$$
n(r)={2\g r\/\pi}+o(r)\qqq\as\qq r\to\iy.
$$
Moreover, for each $\d >0$ the number of zeros of $d(k)$ with
modulus $\leq r$ lying outside both of the two sectors $|\arg k| ,
|\arg k -\pi |<\d$ is $o(r)$ for large $r$.

4) There are only finitely many resonances in the domain $\{k\in
\C_-:|k|>\|p\|e^{\|p\|}e^{-2\g\Im k}\}$.

5)  Lieb-Thirring type inequalities for resonances were determined
in \cite{K12}.

6)
Inverse resonance  problem was solved (characterization,
recovering, plus uniqueness) in terms of resonances for the
Schr\"odinger operator with a compactly supported potential on the
real line \cite{K05} and the half-line \cite{K04},
see also Zworski \cite{Z02}  concerning the uniqueness.

7) Stability estimates  for resonances were determined in
\cite{K04x}, \cite{MSW10}.

 Thus, the problems of resonances for 1-dimensional Schr\"odinger
operators with compactly supported potentials are well understood.

\begin{figure}[t]
\tiny \unitlength 0.7mm \linethickness{0.4pt}
\ifx\plotpoint\undefined\newsavebox{\plotpoint}\fi 
\begin{picture}(85.875,73.5)(0,0)
\put(41.8,71.25){\line(0,-1){70}}
\put(85.125,29.25){\line(-1,0){85}}
\put(14.625,2){\line(1,1){59.5}}
\put(82.875,54){\makebox(0,0)[cc]{$\K_1$}}
\put(10.125,54){\makebox(0,0)[cc]{$\K_2$}}
\put(8.125,3.25){\makebox(0,0)[cc]{$\K_3$}}
\put(80.875,3.25){\makebox(0,0)[cc]{$\K_4$}}
\put(66.375,61.25){\makebox(0,0)[cc]{$e^{i{\pi\/4}}\R$}}
\put(44.625,26.5){\makebox(0,0)[cc]{$0$}}
\put(84.875,32){\makebox(0,0)[cc]{$\Re k$}}
\put(47.625,69.75){\makebox(0,0)[cc]{$\Im k$}}
\qbezier(56.375,29.25)(56.75,19.875)(85.625,18)
\qbezier(41.875,43.75)(32.5,44.125)(30.625,73)
\multiput(32.625,58)(.042600897,.033632287){223}{\line(1,0){.042600897}}
\multiput(70.375,20.25)(.033632287,.042600897){223}{\line(0,1){.042600897}}
\multiput(31.625,64.25)(.043067227,.033613445){238}{\line(1,0){.043067227}}
\multiput(76.625,19.25)(.033613445,.043067227){238}{\line(0,1){.043067227}}
\put(30.625,70.5){\line(4,3){4}}
\put(82.875,18.25){\line(3,4){3}}
\put(67.815,29.25){\line(0,1){1.1155}}
\put(67.791,30.366){\line(0,1){1.1135}}
\put(67.721,31.479){\line(0,1){1.1096}}
\multiput(67.603,32.589)(-.032842,.220723){5}{\line(0,1){.220723}}
\multiput(67.439,33.692)(-.03009,.156529){7}{\line(0,1){.156529}}
\multiput(67.229,34.788)(-.032084,.13573){8}{\line(0,1){.13573}}
\multiput(66.972,35.874)(-.033584,.119338){9}{\line(0,1){.119338}}
\multiput(66.67,36.948)(-.031572,.096394){11}{\line(0,1){.096394}}
\multiput(66.322,38.008)(-.032644,.087062){12}{\line(0,1){.087062}}
\multiput(65.931,39.053)(-.0334964,.0790218){13}{\line(0,1){.0790218}}
\multiput(65.495,40.08)(-.0318938,.0671998){15}{\line(0,1){.0671998}}
\multiput(65.017,41.088)(-.0325318,.0616822){16}{\line(0,1){.0616822}}
\multiput(64.496,42.075)(-.0330403,.0567103){17}{\line(0,1){.0567103}}
\multiput(63.935,43.039)(-.0334366,.0521955){18}{\line(0,1){.0521955}}
\multiput(63.333,43.979)(-.0337348,.0480679){19}{\line(0,1){.0480679}}
\multiput(62.692,44.892)(-.0323297,.0421636){21}{\line(0,1){.0421636}}
\multiput(62.013,45.777)(-.0325307,.0389092){22}{\line(0,1){.0389092}}
\multiput(61.297,46.633)(-.0326589,.0358715){23}{\line(0,1){.0358715}}
\multiput(60.546,47.458)(-.0327206,.0330258){24}{\line(0,1){.0330258}}
\multiput(59.761,48.251)(-.0355668,.0329905){23}{\line(-1,0){.0355668}}
\multiput(58.943,49.01)(-.0386055,.0328906){22}{\line(-1,0){.0386055}}
\multiput(58.093,49.733)(-.0418616,.0327198){21}{\line(-1,0){.0418616}}
\multiput(57.214,50.421)(-.045365,.0324707){20}{\line(-1,0){.045365}}
\multiput(56.307,51.07)(-.0491522,.0321345){19}{\line(-1,0){.0491522}}
\multiput(55.373,51.681)(-.0564011,.0335654){17}{\line(-1,0){.0564011}}
\multiput(54.414,52.251)(-.0613775,.0331031){16}{\line(-1,0){.0613775}}
\multiput(53.432,52.781)(-.0669008,.0325164){15}{\line(-1,0){.0669008}}
\multiput(52.429,53.269)(-.0730855,.0317838){14}{\line(-1,0){.0730855}}
\multiput(51.405,53.714)(-.086755,.033451){12}{\line(-1,0){.086755}}
\multiput(50.364,54.115)(-.096097,.032466){11}{\line(-1,0){.096097}}
\multiput(49.307,54.472)(-.107119,.031221){10}{\line(-1,0){.107119}}
\multiput(48.236,54.784)(-.135427,.033342){8}{\line(-1,0){.135427}}
\multiput(47.153,55.051)(-.156243,.031542){7}{\line(-1,0){.156243}}
\multiput(46.059,55.272)(-.183674,.029075){6}{\line(-1,0){.183674}}
\multiput(44.957,55.446)(-.27711,.03195){4}{\line(-1,0){.27711}}
\put(43.849,55.574){\line(-1,0){1.1128}}
\put(42.736,55.655){\line(-1,0){1.1152}}
\put(41.62,55.689){\line(-1,0){1.1157}}
\put(40.505,55.676){\line(-1,0){1.1141}}
\put(39.391,55.615){\line(-1,0){1.1106}}
\multiput(38.28,55.508)(-.221018,-.030792){5}{\line(-1,0){.221018}}
\multiput(37.175,55.354)(-.182935,-.033408){6}{\line(-1,0){.182935}}
\multiput(36.077,55.154)(-.136022,-.030822){8}{\line(-1,0){.136022}}
\multiput(34.989,54.907)(-.119645,-.032474){9}{\line(-1,0){.119645}}
\multiput(33.912,54.615)(-.096683,-.030676){11}{\line(-1,0){.096683}}
\multiput(32.849,54.278)(-.087361,-.031834){12}{\line(-1,0){.087361}}
\multiput(31.801,53.896)(-.0793294,-.0327613){13}{\line(-1,0){.0793294}}
\multiput(30.769,53.47)(-.0723139,-.033502){14}{\line(-1,0){.0723139}}
\multiput(29.757,53.001)(-.0619816,-.0319577){16}{\line(-1,0){.0619816}}
\multiput(28.765,52.489)(-.0570147,-.0325123){17}{\line(-1,0){.0570147}}
\multiput(27.796,51.937)(-.0525037,-.0329506){18}{\line(-1,0){.0525037}}
\multiput(26.851,51.343)(-.0483791,-.0332871){19}{\line(-1,0){.0483791}}
\multiput(25.932,50.711)(-.044585,-.0335337){20}{\line(-1,0){.044585}}
\multiput(25.04,50.04)(-.0410767,-.0336999){21}{\line(-1,0){.0410767}}
\multiput(24.177,49.333)(-.0361732,-.0323244){23}{\line(-1,0){.0361732}}
\multiput(23.345,48.589)(-.0333282,-.0324125){24}{\line(-1,0){.0333282}}
\multiput(22.545,47.811)(-.0333193,-.0352589){23}{\line(0,-1){.0352589}}
\multiput(21.779,47)(-.0332476,-.0382984){22}{\line(0,-1){.0382984}}
\multiput(21.048,46.158)(-.033107,-.041556){21}{\line(0,-1){.041556}}
\multiput(20.352,45.285)(-.0328905,-.0450616){20}{\line(0,-1){.0450616}}
\multiput(19.695,44.384)(-.0325895,-.0488517){19}{\line(0,-1){.0488517}}
\multiput(19.075,43.456)(-.0321938,-.0529711){18}{\line(0,-1){.0529711}}
\multiput(18.496,42.502)(-.0336716,-.0610675){16}{\line(0,-1){.0610675}}
\multiput(17.957,41.525)(-.0331361,-.066596){15}{\line(0,-1){.066596}}
\multiput(17.46,40.526)(-.032461,-.0727872){14}{\line(0,-1){.0727872}}
\multiput(17.006,39.507)(-.0316197,-.0797913){13}{\line(0,-1){.0797913}}
\multiput(16.595,38.47)(-.033357,-.095791){11}{\line(0,-1){.095791}}
\multiput(16.228,37.416)(-.032214,-.106825){10}{\line(0,-1){.106825}}
\multiput(15.906,36.348)(-.030754,-.120099){9}{\line(0,-1){.120099}}
\multiput(15.629,35.267)(-.032991,-.155943){7}{\line(0,-1){.155943}}
\multiput(15.398,34.175)(-.030779,-.183396){6}{\line(0,-1){.183396}}
\multiput(15.213,33.075)(-.027617,-.221437){5}{\line(0,-1){.221437}}
\put(15.075,31.968){\line(0,-1){1.112}}
\put(14.984,30.856){\line(0,-1){4.4568}}
\multiput(15.089,26.399)(.028738,-.221295){5}{\line(0,-1){.221295}}
\multiput(15.233,25.293)(.031708,-.183238){6}{\line(0,-1){.183238}}
\multiput(15.423,24.193)(.029558,-.136303){8}{\line(0,-1){.136303}}
\multiput(15.66,23.103)(.031362,-.119941){9}{\line(0,-1){.119941}}
\multiput(15.942,22.023)(.032755,-.10666){10}{\line(0,-1){.10666}}
\multiput(16.269,20.957)(.031021,-.087653){12}{\line(0,-1){.087653}}
\multiput(16.642,19.905)(.0320234,-.0796302){13}{\line(0,-1){.0796302}}
\multiput(17.058,18.87)(.0328292,-.0726219){14}{\line(0,-1){.0726219}}
\multiput(17.518,17.853)(.033473,-.0664273){15}{\line(0,-1){.0664273}}
\multiput(18.02,16.856)(.0319815,-.0573141){17}{\line(0,-1){.0573141}}
\multiput(18.563,15.882)(.0324617,-.0528074){18}{\line(0,-1){.0528074}}
\multiput(19.148,14.932)(.0328365,-.048686){19}{\line(0,-1){.048686}}
\multiput(19.772,14.007)(.0331183,-.0448944){20}{\line(0,-1){.0448944}}
\multiput(20.434,13.109)(.033317,-.0413878){21}{\line(0,-1){.0413878}}
\multiput(21.134,12.24)(.0334411,-.0381296){22}{\line(0,-1){.0381296}}
\multiput(21.869,11.401)(.0334974,-.0350897){23}{\line(0,-1){.0350897}}
\multiput(22.64,10.594)(.0349481,-.0336452){23}{\line(1,0){.0349481}}
\multiput(23.443,9.82)(.0379881,-.0336017){22}{\line(1,0){.0379881}}
\multiput(24.279,9.081)(.0412468,-.0334914){21}{\line(1,0){.0412468}}
\multiput(25.145,8.377)(.0447543,-.0333074){20}{\line(1,0){.0447543}}
\multiput(26.04,7.711)(.048547,-.0330417){19}{\line(1,0){.048547}}
\multiput(26.963,7.083)(.0526699,-.0326842){18}{\line(1,0){.0526699}}
\multiput(27.911,6.495)(.0571786,-.0322231){17}{\line(1,0){.0571786}}
\multiput(28.883,5.947)(.0621426,-.0316434){16}{\line(1,0){.0621426}}
\multiput(29.877,5.441)(.0724827,-.0331353){14}{\line(1,0){.0724827}}
\multiput(30.892,4.977)(.0794943,-.0323591){13}{\line(1,0){.0794943}}
\multiput(31.925,4.556)(.087521,-.031391){12}{\line(1,0){.087521}}
\multiput(32.976,4.18)(.106521,-.033205){10}{\line(1,0){.106521}}
\multiput(34.041,3.848)(.119808,-.031868){9}{\line(1,0){.119808}}
\multiput(35.119,3.561)(.136177,-.030133){8}{\line(1,0){.136177}}
\multiput(36.209,3.32)(.183102,-.032481){6}{\line(1,0){.183102}}
\multiput(37.307,3.125)(.221172,-.029672){5}{\line(1,0){.221172}}
\put(38.413,2.976){\line(1,0){1.1111}}
\put(39.524,2.875){\line(1,0){1.1144}}
\put(40.639,2.82){\line(1,0){1.1157}}
\put(41.754,2.813){\line(1,0){1.1151}}
\put(42.869,2.852){\line(1,0){1.1124}}
\multiput(43.982,2.939)(.27694,.03335){4}{\line(1,0){.27694}}
\multiput(45.09,3.072)(.183524,.030005){6}{\line(1,0){.183524}}
\multiput(46.191,3.252)(.156081,.032333){7}{\line(1,0){.156081}}
\multiput(47.283,3.479)(.120227,.030247){9}{\line(1,0){.120227}}
\multiput(48.365,3.751)(.10696,.031763){10}{\line(1,0){.10696}}
\multiput(49.435,4.068)(.095931,.032952){11}{\line(1,0){.095931}}
\multiput(50.49,4.431)(.0799241,.0312826){13}{\line(1,0){.0799241}}
\multiput(51.529,4.838)(.0729236,.0321535){14}{\line(1,0){.0729236}}
\multiput(52.55,5.288)(.0667352,.0328548){15}{\line(1,0){.0667352}}
\multiput(53.551,5.781)(.0612091,.0334135){16}{\line(1,0){.0612091}}
\multiput(54.531,6.315)(.0531065,.03197){18}{\line(1,0){.0531065}}
\multiput(55.486,6.891)(.0489888,.032383){19}{\line(1,0){.0489888}}
\multiput(56.417,7.506)(.0452,.0327){20}{\line(1,0){.0452}}
\multiput(57.321,8.16)(.0416953,.0329313){21}{\line(1,0){.0416953}}
\multiput(58.197,8.851)(.0384384,.0330857){22}{\line(1,0){.0384384}}
\multiput(59.042,9.579)(.0353992,.0331702){23}{\line(1,0){.0353992}}
\multiput(59.857,10.342)(.0325529,.0331911){24}{\line(0,1){.0331911}}
\multiput(60.638,11.139)(.0324768,.0360365){23}{\line(0,1){.0360365}}
\multiput(61.385,11.968)(.0323332,.0390734){22}{\line(0,1){.0390734}}
\multiput(62.096,12.827)(.0337215,.0444431){20}{\line(0,1){.0444431}}
\multiput(62.771,13.716)(.033491,.0482381){19}{\line(0,1){.0482381}}
\multiput(63.407,14.633)(.0331718,.0523642){18}{\line(0,1){.0523642}}
\multiput(64.004,15.575)(.0327526,.0568769){17}{\line(0,1){.0568769}}
\multiput(64.561,16.542)(.032219,.0618462){16}{\line(0,1){.0618462}}
\multiput(65.076,17.532)(.0315531,.0673605){15}{\line(0,1){.0673605}}
\multiput(65.55,18.542)(.0330958,.0791904){13}{\line(0,1){.0791904}}
\multiput(65.98,19.572)(.032202,.087226){12}{\line(0,1){.087226}}
\multiput(66.366,20.618)(.031084,.096553){11}{\line(0,1){.096553}}
\multiput(66.708,21.68)(.032979,.119507){9}{\line(0,1){.119507}}
\multiput(67.005,22.756)(.031396,.135891){8}{\line(0,1){.135891}}
\multiput(67.256,23.843)(.029297,.15668){7}{\line(0,1){.15668}}
\multiput(67.461,24.94)(.031724,.220886){5}{\line(0,1){.220886}}
\put(67.62,26.044){\line(0,1){1.1101}}
\put(67.732,27.154){\line(0,1){2.0956}}
\multiput(36.875,55.25)(.045673077,.033653846){104}{\line(1,0){.045673077}}
\multiput(67.625,24.75)(.033653846,.045673077){104}{\line(0,1){.045673077}}
\multiput(18.875,65.25)(.0333333,-.0333333){30}{\line(0,-1){.0333333}}
\multiput(77.875,6.25)(-.0333333,.0333333){30}{\line(0,1){.0333333}}
\multiput(26.375,59.25)(.0333333,-.0333333){30}{\line(0,-1){.0333333}}
\multiput(71.875,13.75)(-.0333333,.0333333){30}{\line(0,1){.0333333}}
\multiput(23.875,48)(.0333333,-.0333333){30}{\line(0,-1){.0333333}}
\multiput(60.625,11.25)(-.0333333,.0333333){30}{\line(0,1){.0333333}}
\multiput(30.5,61.375)(.0333333,-.0333333){30}{\line(0,-1){.0333333}}
\multiput(74,17.875)(-.0333333,.0333333){30}{\line(0,1){.0333333}}
\multiput(29.25,73.125)(.0333333,-.0333333){30}{\line(0,-1){.0333333}}
\multiput(85.75,16.625)(-.0333333,.0333333){30}{\line(0,1){.0333333}}
\multiput(27.5,64.875)(.0333333,-.0333333){30}{\line(1,0){.0333333}}
\multiput(77.5,14.875)(-.0333333,.0333333){30}{\line(0,1){.0333333}}
\multiput(7.375,20.75)(.0333333,-.0333333){30}{\line(1,0){.0333333}}
\multiput(4.375,15)(.0333333,-.0333333){30}{\line(0,-1){.0333333}}
\multiput(36.75,51.375)(.0333333,-.0333333){30}{\line(0,-1){.0333333}}
\multiput(64,24.125)(-.0333333,.0333333){30}{\line(0,1){.0333333}}
\multiput(51.625,28.875)(-.0333333,.0333333){30}{\line(0,1){.0333333}}
\multiput(41.5,38.75)(.0333333,-.0333333){30}{\line(0,-1){.0333333}}
\multiput(54.125,42.5)(.0333333,-.0333333){30}{\line(0,-1){.0333333}}
\multiput(62.875,51.25)(.0333333,-.0333333){30}{\line(0,-1){.0333333}}
\multiput(70.25,58.375)(.0333333,-.0333333){30}{\line(0,-1){.0333333}}
\multiput(65.5,28.625)(-.0333333,.0333333){30}{\line(0,1){.0333333}}
\multiput(41.625,52.625)(.0328947,-.0526316){19}{\line(0,-1){.0526316}}
\multiput(75.25,28.625)(-.0333333,.0333333){30}{\line(0,1){.0333333}}
\multiput(41.25,62.375)(.0333333,-.0333333){30}{\line(0,-1){.0333333}}
\multiput(14.125,40.625)(.0333333,-.0333333){30}{\line(0,-1){.0333333}}
\multiput(53.25,1.5)(-.0333333,.0333333){30}{\line(0,1){.0333333}}
\multiput(19.125,58.625)(.0333333,-.0333333){30}{\line(0,-1){.0333333}}
\multiput(71.25,6.5)(-.0333333,.0333333){30}{\line(0,1){.0333333}}
\multiput(30.875,39.625)(.0333333,-.0333333){30}{\line(0,-1){.0333333}}
\multiput(52.25,18.25)(-.0333333,.0333333){30}{\line(0,1){.0333333}}
\multiput(24.875,19.25)(.0333333,-.0333333){30}{\line(0,-1){.0333333}}
\multiput(31.875,12.25)(-.0333333,.0333333){30}{\line(0,1){.0333333}}
\multiput(12.375,25.125)(.0333333,-.0333333){30}{\line(0,-1){.0333333}}
\multiput(37.75,-.25)(-.0333333,.0333333){30}{\line(0,1){.0333333}}
\multiput(2.125,18.375)(.0333333,-.0333333){30}{\line(0,-1){.0333333}}
\multiput(23.5,25.5)(.0333333,-.0333333){30}{\line(0,-1){.0333333}}
\multiput(38.125,10.875)(-.0333333,.0333333){30}{\line(0,1){.0333333}}
\multiput(25.25,39.625)(.0333333,-.0333333){30}{\line(0,-1){.0333333}}
\multiput(29.25,29.625)(.0333333,-.0333333){30}{\line(0,-1){.0333333}}
\multiput(41.5,17.125)(.0333333,-.0333333){30}{\line(0,-1){.0333333}}
\multiput(52.25,12.625)(-.0333333,.0333333){30}{\line(0,1){.0333333}}
\multiput(29.75,68.375)(.0333333,-.0333333){30}{\line(0,-1){.0333333}}
\multiput(81,17.125)(-.0333333,.0333333){30}{\line(0,1){.0333333}}
\multiput(19.125,64.5)(.0328947,.0460526){19}{\line(0,1){.0460526}}
\multiput(77.125,6.5)(.0460526,.0328947){19}{\line(1,0){.0460526}}
\multiput(26.625,58.5)(.0328947,.0460526){19}{\line(0,1){.0460526}}
\multiput(71.125,14)(.0460526,.0328947){19}{\line(1,0){.0460526}}
\multiput(24.125,47.25)(.0328947,.0460526){19}{\line(0,1){.0460526}}
\multiput(59.875,11.5)(.0460526,.0328947){19}{\line(1,0){.0460526}}
\multiput(30.75,60.625)(.0328947,.0460526){19}{\line(0,1){.0460526}}
\multiput(73.25,18.125)(.0460526,.0328947){19}{\line(1,0){.0460526}}
\multiput(29.5,72.375)(.0328947,.0460526){19}{\line(0,1){.0460526}}
\multiput(85,16.875)(.0460526,.0328947){19}{\line(1,0){.0460526}}
\multiput(27.75,64.125)(.0328947,.0460526){19}{\line(0,1){.0460526}}
\multiput(76.75,15.125)(.0460526,.0328947){19}{\line(1,0){.0460526}}
\multiput(7.625,20)(.0328947,.0460526){19}{\line(0,1){.0460526}}
\multiput(4.625,14.25)(.0328947,.0460526){19}{\line(0,1){.0460526}}
\multiput(37,50.625)(.0328947,.0460526){19}{\line(0,1){.0460526}}
\multiput(63.25,24.375)(.0460526,.0328947){19}{\line(1,0){.0460526}}
\multiput(50.875,29.125)(.0460526,.0328947){19}{\line(1,0){.0460526}}
\multiput(41.75,38)(.0328947,.0460526){19}{\line(0,1){.0460526}}
\multiput(54.375,41.75)(.0328947,.0460526){19}{\line(0,1){.0460526}}
\multiput(63.125,50.5)(.0328947,.0460526){19}{\line(0,1){.0460526}}
\multiput(70.5,57.625)(.0328947,.0460526){19}{\line(0,1){.0460526}}
\multiput(64.75,28.875)(.0460526,.0328947){19}{\line(1,0){.0460526}}
\multiput(41.5,51.875)(.0328947,.0460526){19}{\line(0,1){.0460526}}
\multiput(74.5,28.875)(.0460526,.0328947){19}{\line(1,0){.0460526}}
\multiput(41.5,61.625)(.0328947,.0460526){19}{\line(0,1){.0460526}}
\multiput(14.375,39.875)(.0328947,.0460526){19}{\line(0,1){.0460526}}
\multiput(52.5,1.75)(.0460526,.0328947){19}{\line(1,0){.0460526}}
\multiput(19.375,57.875)(.0328947,.0460526){19}{\line(0,1){.0460526}}
\multiput(70.5,6.75)(.0460526,.0328947){19}{\line(1,0){.0460526}}
\multiput(31.125,38.875)(.0328947,.0460526){19}{\line(0,1){.0460526}}
\multiput(51.5,18.5)(.0460526,.0328947){19}{\line(1,0){.0460526}}
\multiput(25.125,18.5)(.0328947,.0460526){19}{\line(0,1){.0460526}}
\multiput(31.125,12.5)(.0460526,.0328947){19}{\line(1,0){.0460526}}
\multiput(12.625,24.375)(.0328947,.0460526){19}{\line(0,1){.0460526}}
\multiput(37,0)(.0460526,.0328947){19}{\line(1,0){.0460526}}
\multiput(2.375,17.625)(.0328947,.0460526){19}{\line(0,1){.0460526}}
\multiput(23.75,24.75)(.0328947,.0460526){19}{\line(0,1){.0460526}}
\multiput(37.375,11.125)(.0460526,.0328947){19}{\line(1,0){.0460526}}
\multiput(25.5,38.875)(.0328947,.0460526){19}{\line(0,1){.0460526}}
\multiput(29.5,28.875)(.0328947,.0460526){19}{\line(0,1){.0460526}}
\multiput(41.75,16.375)(.0328947,.0460526){19}{\line(0,1){.0460526}}
\multiput(51.5,12.875)(.0460526,.0328947){19}{\line(1,0){.0460526}}
\multiput(30,67.625)(.0328947,.0460526){19}{\line(0,1){.0460526}}
\multiput(80.25,17.375)(.0460526,.0328947){19}{\line(1,0){.0460526}}
\end{picture}\caption{\footnotesize The plane of variable $k$. The function
$D(k)$ is real on the line $e^{i{\pi\/4}}\R$. The resonances are marked by crosses.
The forbidden domain
for the resonances is shaded.}
\lb{FigPk}
\end{figure}
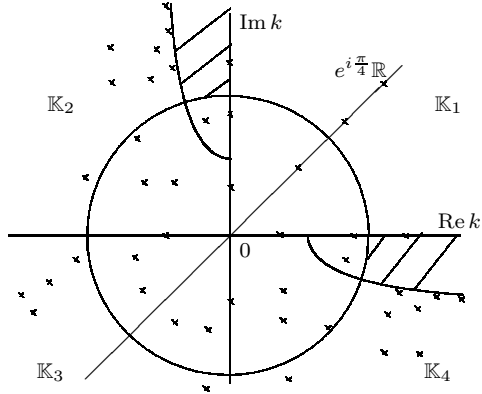

\subsection{Determinant}
 Instead
of the spectral parameter $\l\in \C$ we introduce a new variable
$k=\l^{1\/4}\in \K_1$, where $\K_1$ is the quadrant given by
$$
\K_1=\big\{k\in\C: \arg k\in
(0,\tf{\pi}{2})\big\}.
$$
The mapping $k=\l^{1\/4}\in \C$ gives a parametrization of the four
sheeted Riemann surface $\L$ of the function $\l^{1\/4}$ as a
complex plane of the variable $k$. The sheet $\L_j=\C\sm
\R_+,j=1,2,3,4$ of the surface $\L$ corresponds to the quadrant
$$
\K_j=i^{j-1}\K_1,\qqq j=1,2,3,4,
$$
of the complex plane of the variable $k$, see Fig.~\ref{FigPk}. The
first sheet $\L_1$ is physical, while the other sheets
$\L_2,\L_3,\L_4$ are non-physical.

In order to define the Fredholm determinant we  rewrite
the perturbation $V$ in the form
\[
\lb{defVj}
\begin{aligned}
V=V_1V_2,\qqq V_1=(\pa |2p|^{1\/2}, |q|^{1\/2}),\qqq  V_2=\ma
(2p)^{1\/2}\pa \\ q^{1\/2}\am,\qqq p^{1\/2}=|p|^{1\/2}\sign p.
\end{aligned}
\]
We set
\[
\lb{Y0}
 R_0(k)=(H_0-k^4)^{-1},\qq  Y_0(k)=V_2 R_0(k)V_1,\qq
 k \in \K_1.
\]
In the next proposition we show that  each operator $Y_0(k),k\in
\K_1,$ is trace class and is analytic in the plane without zero.
Thus we can define the Fredholm determinant $D(k)$ by
\[
\label{a.2}
D(k)=\det (I +Y_0(k)),\qqq k\in \C\sm \{0\}.
\]
The function $D$  is analytic in $\C\sm \{0\}$. Note that
$k\in\ol\K_1\sm\{0\}$ is a zero of the determinant $D$ iff
$\l=k^4\in \R\sm\{0\}$ is an eigenvalue of the operator $H$. We
define the resonances as zeros of a Fredholm determinant in $\C\sm
\ol \K_1$. We formulate our preliminary results about the
determinant.

\begin{proposition}
\lb{T1}

Let $p,q\in\cH_0$. Then

i) Each operator $Y_0(k),k\in \K_1$, is trace class  and the
operator-valued function $kY_0(k)$ is entire.

ii) The Fredholm determinant $D(k)$  is analytic
in $\K_1$ and has an analytic extension into the whole plane without
zero such that the function $kD(k)$ is entire.
 In particular, the operator $H$ has a finite number of eigenvalues.
 Moreover, $D(k)$ is real on the line $e^{i{\pi\/4}}\R$ and satisfies:
\[
\lb{sym}
D(k)=\ol{D(i\ol k)}\qqq \forall \ k\in \C\sm\{0\},
\]
\[
\lb{asDK+}
D(k)=1-\fr{1+i}{2k}\int_{\R_+} p(x)dx+\fr{O(1)}{k^{2}}\qqq
\as \qqq |k| \to \infty, \qq  k\in\ol\K_1,
\]
uniformly in $\arg k\in[0,{\pi\/2}]$.

\end{proposition}

\no {\bf Remark.}
Due to \er{sym} the function $D(k)$ is symmetric with respect to
the line $e^{i{\pi\/4}}\R$. Thus it is sufficiently
to analyze this function in the half-plane $e^{i{\pi\/4}}\ol\C_+$.

\medskip

Recall that the zeros of the function $D$ in $\C\sm\ol\K_1$ are
called {\it resonances} of $H$. Let $\cN(r)$ be the number of zeros
of the function $D$ in the disc $|k|<r$, counted with multiplicity.
We present our first main result.

\begin{theorem}
\lb{CorEstN}
Let $p,q\in\cH_0$ and let $k_*\in\K_2$ be a resonance.
Then
\[
\lb{unifestD}
|D(k)|\le Ce^{2\g( (\Re k)_-+(\Im
k)_-)},\qqq\forall\qq k\in\C,\qq |k|\ge 1,
\]
\[
\lb{estnr} \cN(r)\le \fr{4\g r}{\pi}\big(1+o(1)\big)\qqq\as\qq
r\to\iy,
\]
\[
\lb{estFD}
|k_*|\le Ce^{-2\g\Re k_*},
\]
for some constant $C=C(p,q)>0$, where $(a)_-={|a|-a\/2}$.
\end{theorem}

\no {\bf Remark.}
1)  Due to \er{estFD} the domain
$\{k\in\K_2:|k|>Ce^{-2\g\Re k}\}$ is forbidden for the resonances
in $\K_2$ and by the symmetry \er{sym} of the function $D(k)$,
the domain $\{k\in\K_4:|k|>Ce^{-2\g\Im k}\}$ is also forbidden
for the resonances in $\K_4$,
see Fig.~\ref{FigPk}.

2) Estimate \er{unifestD} is crucial to prove trace formulas
in terms of resonances in Theorem~\ref{ThRes}.

 3) Consider a specific case when the  coefficients $p,q$ satisfy $q=p''+p^2$. Then
$H$ has the form $H=h^2$, where $h$ is the Scr\"odinger operator,
given by \er{Schr}.
The determinant $D$ for the operator $h^2$
satisfies the identity
$$
D(k)=d(ik)d(k)\qqq \forall\qq k\in\C,
$$
see \cite{BK16}, where $d(k)$ is the determinant for the operator
$h$, given by \er{dt2}. Thus in this case we can describe resonances
of $H$, see Fig.~\ref{fighsq}~b).
For example, due to Zworski
\cite{Z87}, we obtain the asymptotic
distribution of resonances.
The resonance of the operator $h$, and then of $h^2$,
may have any multiplicity, see \cite{K04}.
Moreover, we can
determine the forbidden domains for the resonances of the operator
$H=h^2$ in all quadrants $\K_2,\K_3,\K_4$ (on all non-physical
sheets $\L_2,\L_3,\L_4$).

\begin{figure}[t]
\tiny
\unitlength 0.6mm
\linethickness{0.4pt}
\ifx\plotpoint\undefined
\newsavebox{\plotpoint}\fi 
\begin{picture}(258.03,105.88)(10,10)
\put(15.21,62.18){\line(1,0){117.32}}
\put(145.71,62.18){\line(1,0){117.32}}
\put(71.21,113.7){\line(0,-1){99.12}}
\put(201.71,113.7){\line(0,-1){99.12}}
\put(25.01,18.5){\makebox(0,0)[cc]{$a)$}}
\put(155.51,18.5){\makebox(0,0)[cc]{$b)$}}
\put(95.01,53.22){\circle*{1.96}}
\put(225.51,53.22){\circle*{1.96}}
\put(192.97,38.24){\circle*{1.96}}
\put(47.41,53.22){\circle*{1.96}}
\put(177.91,53.22){\circle*{1.96}}
\put(192.97,85.84){\circle*{1.96}}
\put(107.05,45.94){\circle*{1.96}}
\put(237.55,45.94){\circle*{1.96}}
\put(185.69,26.2){\circle*{1.96}}
\put(35.37,45.94){\circle*{1.96}}
\put(165.87,45.94){\circle*{1.96}}
\put(185.69,97.88){\circle*{1.96}}
\put(108.73,53.78){\circle*{1.96}}
\put(239.23,53.78){\circle*{1.96}}
\put(193.53,24.52){\circle*{1.96}}
\put(33.69,53.78){\circle*{1.96}}
\put(164.19,53.78){\circle*{1.96}}
\put(193.53,99.56){\circle*{1.96}}
\put(116.57,45.66){\circle*{1.96}}
\put(247.07,45.66){\circle*{1.96}}
\put(185.41,16.68){\circle*{1.96}}
\put(25.85,45.66){\circle*{1.96}}
\put(156.35,45.66){\circle*{1.96}}
\put(185.41,107.4){\circle*{1.96}}
\put(71.21,36.14){\circle*{1.96}}
\put(201.71,36.14){\circle*{1.96}}
\put(175.89,62.04){\circle*{1.96}}
\put(71.21,29.98){\circle*{1.96}}
\put(201.71,29.98){\circle*{1.96}}
\put(169.73,62.04){\circle*{1.96}}
\put(71.21,40.06){\circle*{1.96}}
\put(201.71,40.06){\circle*{1.96}}
\put(179.81,62.04){\circle*{1.96}}
\put(71.21,20.74){\circle*{1.96}}
\put(201.71,20.74){\circle*{1.96}}
\put(160.49,62.04){\circle*{1.96}}
\put(78.21,55.46){\circle*{1.96}}
\put(208.71,55.46){\circle*{1.96}}
\put(195.21,55.04){\circle*{1.96}}
\put(64.21,55.46){\circle*{1.96}}
\put(194.71,55.46){\circle*{1.96}}
\put(195.21,69.04){\circle*{1.96}}
\put(78.21,47.62){\circle*{1.96}}
\put(208.71,47.62){\circle*{1.96}}
\put(187.37,55.04){\circle*{1.96}}
\put(64.21,47.62){\circle*{1.96}}
\put(194.71,47.62){\circle*{1.96}}
\put(187.37,69.04){\circle*{1.96}}
\put(103.97,57.7){\circle*{1.96}}
\put(234.47,57.7){\circle*{1.96}}
\put(197.45,29.28){\circle*{1.96}}
\put(38.45,57.7){\circle*{1.96}}
\put(168.95,57.7){\circle*{1.96}}
\put(197.45,94.8){\circle*{1.96}}
\put(125.53,45.66){\circle*{1.96}}
\put(256.03,45.66){\circle*{1.96}}
\put(16.89,45.66){\circle*{1.96}}
\put(147.39,45.66){\circle*{1.96}}
\put(185.41,116.36){\circle*{1.96}}
\put(130.85,43.42){\circle*{1.96}}
\put(261.35,43.42){\circle*{1.96}}
\put(89.69,59.66){\circle*{1.96}}
\put(220.19,59.66){\circle*{1.96}}
\put(199.41,43.56){\circle*{1.96}}
\put(52.73,59.66){\circle*{1.96}}
\put(183.23,59.66){\circle*{1.96}}
\put(199.41,80.52){\circle*{1.96}}
\put(71.21,83.18){\circle*{1.96}}
\put(201.71,83.18){\circle*{1.96}}
\put(222.93,62.04){\circle*{1.96}}
\put(71.21,101.66){\circle*{1.96}}
\put(201.71,101.66){\circle*{1.96}}
\put(241.41,62.04){\circle*{1.96}}
\put(71.21,111.74){\circle*{1.96}}
\put(201.71,111.74){\circle*{1.96}}
\put(251.49,62.04){\circle*{1.96}}
\put(131.97,64.7){\makebox(0,0)[cc]{$\Re k$}}
\put(262.47,64.7){\makebox(0,0)[cc]{$\Re k$}}
\put(64.21,108.94){\makebox(0,0)[cc]{$\Im k$}}
\put(211.21,108.94){\makebox(0,0)[cc]{$\Im k$}}
\qbezier(36.77,61.9)(36.77,55.74)(14.37,50.7)
\qbezier(167.27,61.9)(167.27,55.74)(144.87,50.7)
\qbezier(201.65,96.48)(195.49,96.48)(190.45,118.88)
\qbezier(105.65,61.9)(105.65,55.74)(128.05,50.7)
\qbezier(236.15,61.9)(236.15,55.74)(258.55,50.7)
\qbezier(201.65,27.6)(197.49,27.6)(193.,15.2)
\put(24,62.12){\line(-1,-1){9}}
\put(154.5,62.12){\line(-1,-1){9}}
\put(201.87,109.25){\line(-1,1){9}}
\put(32,62.12){\line(-1,-1){9.5}}
\put(162.5,62.12){\line(-1,-1){9.5}}
\put(201.87,101.25){\line(-1,1){9.5}}
\put(117,62.12){\line(-1,-1){6}}
\put(247.5,62.12){\line(-1,-1){6}}
\put(201.87,16.25){\line(-1,1){6}}
\put(124.5,62.12){\line(-1,-1){8}}
\put(255,62.12){\line(-1,-1){8}}
\put(132,62.12){\line(-1,-1){10}}
\put(262.5,62.12){\line(-1,-1){10}}
\end{picture}
\caption{\footnotesize a) Resonances for the second order operator
$h$; b) Resonances for the fourth order operator $h^2$.
The resonances are marked by circles.
The forbidden domains for the resonances are shaded.
}
\lb{fighsq}
\end{figure}
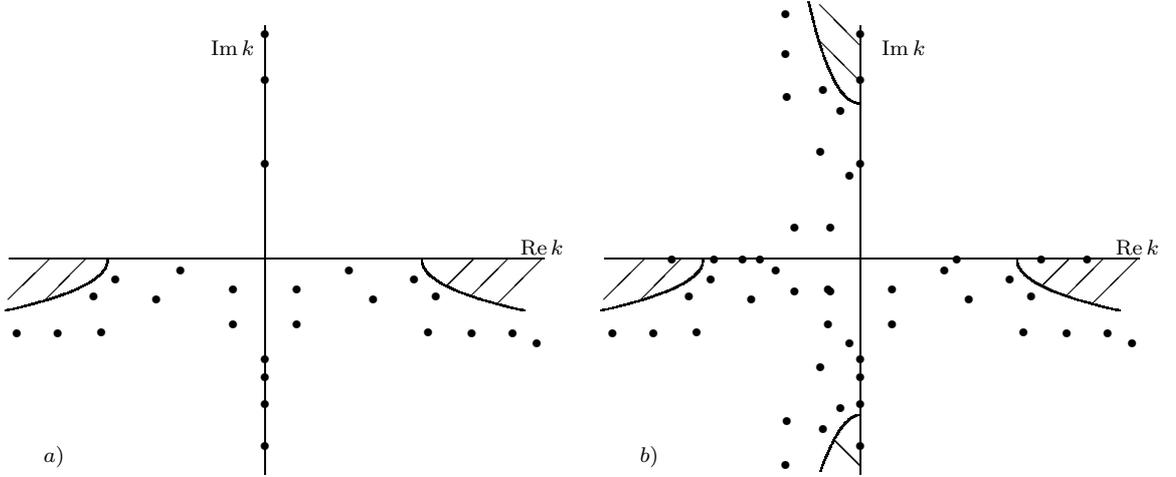

\subsection{Asymptotics of resonances}
In order to determine asymptotics of resonances we assume the
stronger conditions for the coefficients $p,q$ given by
\[
\lb{pqp}
q\in \cH_0,\qq p\in \cH_1 \qqq {\rm and } \qq p_+:=p(\g-0)\ne 0.
\]
 Introduce the
numbers $k_{\pm n}^0, n\in \N$ by
\[
\lb{defzn}
\begin{aligned}
k_{n}^0={1\/\g}\rt(i\pi j_n-\log {2\pi n\/\g |2p_+|^{1\/2}}\rt),\qq
k_{-n}^0=ik_{n}^0-{\pi\/2\g },\qq n\in\N,
\end{aligned}
\]
where $j_n=\ca n,& \text{if}\ p_+>0\\n+{1\/2},& \text{if}\
p_+<0\ac$, see Fig.~\ref{figas}.

\begin{theorem}
\lb{ThAsCF} Let $p,q$ satisfy \er{pqp}. Then for any $\ve\in
(0,{\pi\/2\g})$  there exists $\r>0$ such that in each disk
$\{|k-k_n^0|<\ve\}\ss e^{i{\pi\/4}}\C_+\cap \{|k|>\r\}$ there exists
exactly one resonance $k_n$
 and there are no other
resonances in the domain $e^{i{\pi\/4}}\C_+\cap \{|k|>\r\}$.
These resonances satisfy
\[
\lb{asresiK+}
k_{n}=k_{n}^0+o(1)\qqq as\qq n\to \pm\iy.
\]
In particular, there is a finite number of resonances on $\R\cup i\R$.

Let $\cN_j(r), j=2,3,4$ be  the number of zeros of the
function $D$ in a domain $\K_j\cap \{|k|<r\}$ counted with
multiplicity, where $\K_j=i^{j-1}\K_1$ and let $ r\to\iy$. Then we
have
\[
\lb{asNj}
\begin{aligned}
\cN_2(r)=\cN_4(r)=\fr{\g r}{\pi}\big(1+o(1)\big),\\
\cN_3(r)=\fr{2\g r}{\pi}\big(1+o(1)\big).
\end{aligned}
\]
\end{theorem}

\no {\bf Remark.} 1) Due to the identity \er{sym},
for each resonance in the domain $e^{i{\pi\/4}}\C_+$ there
exists the symmetric resonance in $e^{i{\pi\/4}}\C_-$ with the same
multiplicity.
Thus Theorem \ref{ThAsCF} describes all resonances outside the large disc.

2)  Roughly speaking
the proof of Theorem~\ref{ThAsCF} is based on the fact
that the Born approximation is the ``main term''
of the scattering amplitude at large $k$.

3) From \er{asNj} we obtain  $\cN_3(r)=2\cN_2(r)(1+o(1))$. This
asymptotics shows the distribution of the resonances on the Riemann
surface $\L$ of the function $\l^{1\/4}$: the number of resonances
in the large disc on the sheet $\L_3$ is in two times more than on
the sheet $\L_2$ (and $\L_4$). Note that the corresponding question
for a second order operator has no meaning, since the Riemann
surface for this case has only one non-physical sheet.

\begin{figure}[t]
\tiny
\unitlength 0.6mm 
\linethickness{0.4pt}
\ifx\plotpoint\undefined\newsavebox{\plotpoint}\fi 
\begin{picture}(137.25,137.75)(0,0)
\put(68.75,133.25){\line(0,-1){132.25}}
\put(2,69.5){\line(1,0){133}}
\put(15,15.5){\line(1,1){105}}
\bezier{40}(47.5,137.75)(47.75,112.625)(54,97.75)
\bezier{30}(54,97.75)(58,86.625)(54,70)
\bezier{50}(54,70)(48.75,48.75)(.25,47.75)
\bezier{40}(137.25,48)(112.125,48.25)(97.5,55)
\bezier{30}(97.5,55)(86.125,58.5)(69.5,55)
\bezier{50}(69.5,55)(48.25,49.25)(47.25,.75)
\put(92.25,57){\circle*{1.5}}
\put(56.5,92.75){\circle*{1.5}}
\put(45.25,57){\circle*{1.5}}
\put(56.5,45.75){\circle*{1.5}}
\put(111.25,50.25){\circle*{1.5}}
\put(49.75,111.75){\circle*{1.5}}
\put(26.25,50.25){\circle*{1.5}}
\put(49.75,26.75){\circle*{1.5}}
\put(130.5,47.25){\circle*{1.5}}
\put(46.75,131){\circle*{1.5}}
\put(7,47.25){\circle*{1.5}}
\put(46.75,7.5){\circle*{1.5}}
\put(89,70.5){\line(0,-1){2}}
\put(110,70.5){\line(0,-1){2}}
\put(130.25,70.75){\line(0,-1){2}}
\put(88,76){\makebox(0,0)[cc]{${\pi\/\g}$}}
\put(109.75,76){\makebox(0,0)[cc]{${2\pi\/\g}$}}
\put(129,76){\makebox(0,0)[cc]{${3\pi\/\g}$}}
\put(133.75,63.75){\makebox(0,0)[cc]{$\Re k$}}
\put(75.5,126.25){\makebox(0,0)[cc]{$\Im k$}}
\put(72,67){\makebox(0,0)[cc]{$0$}}
\end{picture}
\caption{\footnotesize Asymptotics of resonances for step
coefficients}
\lb{figas}
\end{figure}
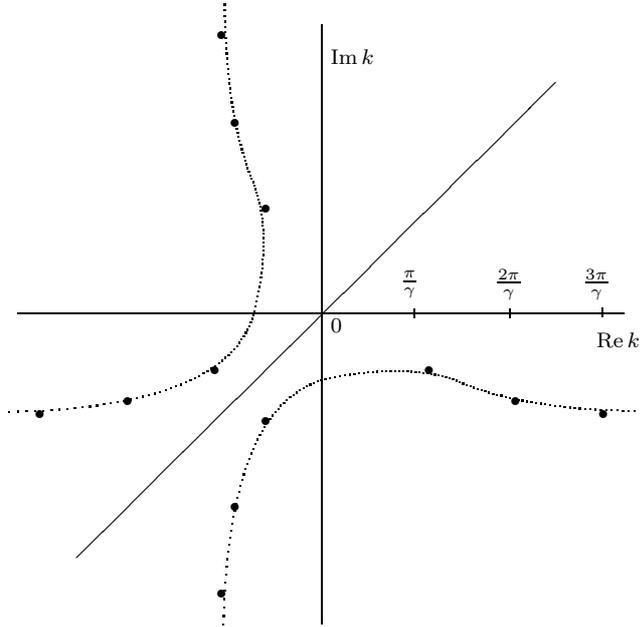

\subsection{Euler-Bernoulli operators}
We consider an Euler-Bernoulli operator $\cE\ge 0$ acting on
$L^2(\R_+,b(x)dx)$ and given by
$$
\cE u={1\/b}(au'')'',
$$
with the boundary conditions
\[
\lb{ebdc1}
u=0,\qqq \text{and} \qq u''+{a'\/5a}u'=0\qq
\text{at}\qq x=0.
\]
We assume that the coefficients $a,b$ are positive on the unit
interval $[0,1]$ and $a=b=1$ outside this interval and satisfy
\[
\lb{abc1}
a-1,b-1\in\cH_4,\qq
\Big({3a'\/a}+{5b'\/b}\Big)\Big|_{x=0}=0.
\]

The Euler-Bernoulli operator describes the relationship between the
thin beam's deflection and the applied load, $a$ is the rigidity and
$b$ is the density of the beam, see, e.g., \cite{TW59}. The boundary
conditions \er{ebdc1} mean that the end of the beam is restrained by
some special rotational spring device.

The standard Liouville type transformation (see Sect \ref{sect55})
yields that the operator $\cE\ge 0$ with the boundary conditions
\er{ebdc1}  is unitarily equivalent to a
fourth order operator $H\ge 0$ with the  boundary
conditions \er{4g.dc} and with specific coefficients $p,q\in
\cH_0$. Using this transformation we can define the
determinant for the Euler-Bernoulli operator $\cE$. The resonances
for the operator $\cE$ coincide with the resonances for the operator
$H$. Thus all results for the resonances of the operator $H$ can be
carried over the operator $\cE$. In particular, we have the
following corollary of Theorem~\ref{CorEstN}.

\begin{corollary}
\lb{CorEB}
Let the coefficients  $a,b$  satisfy  the conditions \er{abc1}.
Then the determinant $D(k)$,
the counting function $\cN(r)$ and the resonances for the Euler-Bernoulli operator $\cE$
satisfy the estimates \er{unifestD}--\er{estFD}, where
\[
\lb{gammaEB}
\g=\int_0^1\Big({b(x)\/a(x)}\Big)^{1\/4}dx.
\]
\end{corollary}

\no {\bf Remark.} There is an interesting problem to study
resonances for the Euler-Bernoulli operator
under the condition $a-1,b-1\in \cH_1$.

\medskip

We recall the Borg type uniqueness result for the Euler-Bernoulli
operator on a finite interval. We consider the
operator $ \bE={1\/b}(au'')'' $ on the interval $[0,1]$
with the boundary conditions
$$ u(0)=u(1)=u''(0)=u''(1)=0.
$$
We assume that the coefficients $a,b$ are positive and  satisfy:
$$
 a,a'''',b,b''''\in L^1(0,1),\qq
\int_0^1\Big({b\/a}\Big)^{1\/4}dx=1,\qq a'(0)=a'(1)=b'(0)=b'(1)=0.
$$
It was proved in \cite{BK15} that the eigenvalues of the operator $\bE$ are $(\pi n)^4$
for all $n\ge 1$ iff $a=b=1$.

Now we consider a Borg type result for resonance scattering
of the Euler-Bernoulli operator $\cE$ on the half-line.

\begin{theorem}
\lb{ThEB} Let the coefficients  $a,b$  satisfy the conditions
\er{abc1}. Then the operator $\cE$ does not have any eigenvalues and
resonances iff $a=b=1$ on the whole half line.
\end{theorem}

\no {\bf Remark.}  Assume that $a'(0)=b'(0)=0$. Then  the
boundary conditions \er{ebdc1} take the form of the conditions for a
pinned beam: $ u(0)=u''(0)=0. $ Moreover, in this case the last
condition in \er{abc1} also holds true.

\subsection{Historical review}

  A lot of papers are
devoted to the direct and inverse spectral problems for fourth order
operators on the line: Aktosun and Papanicolaou \cite{AP08}, Butler
\cite{B68}, Beals \cite{B85}, Iwasaki \cite{I88}, \cite{I88x},
Hoppe, Laptev and \"Ostensson \cite{HLO06}, Laptev, Shterenberg,
Sukhanov and \"Ostensson \cite{LSSO06}. Moreover, there is a paper
\cite{B85} and even a book \cite{BDT88} about scattering for
1-dimensional higher order operators. However, even the inverse
scattering problems for fourth order operators on the line (or
half-line) are not solved and there no results about resonances.

Eigenvalue asymptotics for fourth order operators
and for the Euler-Bernoulli operators on the finite interval
were determined by Badanin and Korotyaev \cite{BK15}.
Eigenvalue asymptotics for fourth order operators on the circle
were the subject of our paper \cite{BK14},
see also Mikhailets and Molyboga \cite{MM12}
for the case of distribution coefficients.

We recall that resonances, from a physicists point of view,
were first studied by Gamov \cite{Ga}.
Since then, properties of  resonances have been the
object of intense study and we refer to   \cite{SZ91} for the
mathematical approach in the multi-dimensional case  and references
given therein.
We discuss the resonances for one-dimensional systems
and higher order operators.
The properties of higher order operators are very different from the
properties of second order systems. In particular, all fundamental
solutions of a second order system are bounded on the real axis,
while a higher order equation has always exponentially growing
solutions. Moreover, the Riemann surface for higher order operators
is more complicated, than the Riemann surface for Schr\"odinger
operator with matrix-valued potentials. Thus the study of higher
order operators require substantial modifications of methods used in
the study of matrix second order operators, see, e.g., \cite{I88},
\cite{I88x}. Nedelec \cite{N07} considered the resonances for
Schr\"odinger operator with matrix-valued compactly supported
potentials on the line. The resonance scattering for third order
operators on the line was considered in \cite{K16}. It is a first
paper on the resonances for higher order operators. Resonance are
defined as zeros of a Fredholm determinant on a non-physical sheet
of three sheeted Riemann surface. Here  upper bounds of the number
of resonances in complex discs at large radius  and  the trace
formula in terms of resonances only are obtained.
The asymptotics of  counting function for resonances is still open
problem, since the standard Born approximation
is not a main term in the high energy asymptotics for the scattering
amplitude, as we
have for second and fourth order operators. In the present paper we
use different methods from \cite{K16}. Note that the situation for a
fourth order operator is simpler, than for a third one, because the
Born term helps us to obtain asymptotics of counting function for
resonances at large radius.

\section {\lb{Sec2} Properties of the free resolvent}
\setcounter{equation}{0}

\subsection{The well-known facts}
By $\cB$ we denote the class of bounded operators.
Let $\cB_1$ and $\cB_2$ be the trace and
the Hilbert-Schmidt class equipped with the norm $\|\cdot \|_{\cB_1}$
and $ \|\cdot \|_{\cB_2}$ correspondingly.
We recall some well known facts. Let $A, B\in \cB$ and $AB, BA,X\in\cB_1$. Then
\[
\lb{2.1}
\Tr AB=\Tr BA,
\]
\[
\lb{2.2}
\det (I+ AB)=\det (I+BA),
\]
\[
\lb{B10}
{\rm the\ mapping}\qq X\to \det (I+ X) \qq {\rm is\  continuous\ on}\ \cB_1,
\]
\[
\lb{B1}
|\det (I+ X)|\le e^{\|X\|_{\cB_1}},
\]
for all $X\in \cB_1$,
see e.g., Sect. 3. in the book \cite{S05}.
Let  the operator-valued function $X :\cD\to \cB_1$ be analytic for
some domain $\cD\ss\C$ and $(I+X (z))^{-1}\in \cB$ for any $z\in
\cD$. Then for the function $F(z)=\det (I+X (z))$ we have
\[
\lb{2.3}
F'(z)= F(z)\Tr (I+X (z))^{-1}X '(z),\qqq \ z\in \cD.
\]

Introduce the space $L^p(\R_+)$ equipped with the norm
$$
\|f\|_p=\Big(\int_{\R_+}|f(x)|^p dx\Big)^{1\/p}\ge 0,\qqq {\rm and\
let} \qqq \|f\|=\|f\|_2.
$$

\subsection{Schr\"odinger operators}
\lb{sSect2nd}
Let the Schr\"odinger operator $h$
and the unperturbed operator $h_0$ be defined by \er{Schr}.

$\bu$
The free resolvent
$r_0(k)=(h_0-k^2)^{-1},k\in\C_+$
is an integral operator
having the kernel $r_0(x,y,k),x,y\in \R_+$
given by
\[
\lb{h1}
r_0(x,y,k)={i\/2k}\Big(e^{ik|x-y|}-e^{ik(x+y)} \Big),\qq k\in\C_+.
\]

$\bu$ Define the operator-valued function $g(k)=\a r_0(k)\b, k\in \C_+,$
for some $\a^2,\b^2\in\cH_0$. For each $k\in \C_+$ the operator
$g(k)\in\cB_j,j=1,2$ and the mapping
\[
\lb{ga}
g(k): \C_+\to \cB_j
\]
is analytic and it has an analytic extension into whole complex
plane without zero. Thus the operator-valued function $g:\C\sm\{0\}\to\cB_1$
is analytic.
Moreover, we have the following estimate
$$
\|g(k)\|_{\cB_2}\le{\|\a\|\|\b\|\/|k|},\qqq\forall\qq k\in \ol\C_+\sm\{0\}.
$$
Define the Fourier transformation $\F_s: L^2(\R_+)\to L^2(\R_+)$ by
\[
\lb{Ft}
\wt f(\x)=(\F_s f)(\x)=\sqrt{2\/\pi}\int_0^\iy f(x)\sin \x xdx,\qqq \x\in\R_+.
\]
Then $r_0(k)=\F_s^* \e_{-k}\e_k\F_s$, where $\e_k(\x)$
is the multiplication by
$(\x-k)^{-1}$ and we have
\[
\lb{arb1}
\begin{aligned}
\|g(k)\|_{\cB_1}=\big\|\a (h_0-k^2)^{-1}\b\big\|_{\cB_1}
\le\big\|\a \F_s^* \e_{-k}\big\|_{\cB_2}
\big\|\e_{k}\F_s\b \big\|_{\cB_2}
\\
\le
{2\|\a\|\|\b\|\/\pi}\|\e_{-k}\|\|\e_{k}\|\le
{2\|\a\|\|\b\|\/\Im k},\qqq \Im k>0,
\end{aligned}
\]
since $\int_0^\iy|\x\pm k|^{-2}d\x\le{\pi\/\Im k}$.

$\bu$ The Schr\"odinger equation
$-y''-py=k^2 y,k\in\C\sm\{0\},p\in\cH_0$, has a unique Jost solution
$f_+(x,k)$ satisfying the condition $
f_+(x,k)=e^{ikx},x>\g$.
The Jost function $f_+(0,k)$ is entire and
satisfies
\[
\lb{aspk}
f_+(0,k)=1+o(1)\qqq  \as \qq |k|\to\iy,
\]
uniformly in $\arg k\in [0,\pi].$
The function $f_+(0,k)$ has a finite number of simple zeros
$i\vk_1,...,i\vk_N$ in $\C_+$ and
the functions $f_+(x,i\vk_n),n=1,...,N$,
are eigenfunctions of $h$ corresponding
to the eigenvalues $-\vk_n^2$.
Moreover, $f_+(0,k)$
has an infinite number of zeros (resonances) in $\C_-$.
There are not any zeros on the real axis with only exception
$k=0$, where the function $f_+(0,k)$ may have a simple zero.

The determinant $d(k)$, given by the definition \er{dt2}, satisfies
the identity $d(k)=f_+(0,k) $ for all $k\in\C$. The scattering
matrix $s(k)$ for the pair $h,h_0$ has the form
\[
\lb{cS(k)2o} s(k)={d(-k)\/d(k)},\qqq k>0.
\]
Using \er{aspk} we can define the function $\log d(k)$ for large
$|k|,k\in\C_+$, by $\log d(k)=o(1)$ as $\Im k\to\iy$. It satisfies
$$
i\log d(k)={1\/2k}\Big(\int_{\R_+}p(x)dx+o(1)\Big) \qq \as \qq \Im k\to+\iy.
$$

\subsection{The free resolvent}
The free resolvent $R_0(k)=(H_0-k^4)^{-1}, k\in\K_1$  has a
representation in terms of the resolvent $r_0$ for $h_0$:
\[
\lb{R0r0}
R_0(k)={r_0(k)-r_0(ik)\/2k^2},
\qqq
\pa R_0(k)\pa=-{r_0(k)+r_0(ik)\/2}.
\]
Then, by \er{h1}, the operator $R_0(k)$ is an integral operator with
the kernel $R_0(x,y,k)$ given by
\[
\lb{R01}
 R_0(x,y,k)={i\/4k^3}\Big(e^{ik|x-y|}
 -e^{ik(x+y)}+ie^{-k|x-y|}-ie^{-k(x+y)} \Big),\qq x,y>0.
\]
For each $x,y\in\R_+$ the function $R_0(x,y,\cdot)$ is analytic in $\C\sm\{0\}$
and has a simple pole at $k=0$:
\[
\lb{R0at0}
R_0(x,y,k)=
{(1+i)\/4k}xy+{(i-1)\/24}\big((x+y)^3-|x-y|^3\big)+O(k)
\]
as $k\in\C,|k|\to 0$ uniformly in bounded $x,y\in \R_+$.
Let $\a^2,\b^2\in\cH_0$. The last asymptotics shows that
each function $\a(x)\big(R_0(x,y,k)-{(1+i)\/4k}xy\big)\b(y),x,y\in\R_+$
is entire in $k$ and the function
${(1+i)\/4k}x\a(x)y\b(y)$ is a kernel
of the rank one operator.

\medskip

Define the operator-valued function
$G(k)=\a \pa R_0(k)\b, k\in \K_1,$
where $\a^2,\b^2\in\cH_0$.
The identity \er{R0r0} yields that
for each $k\in \K_1$ the operator
$G(k)\in\cB_j,j=1,2$ and the mappings
\[
\lb{Ga}
G(k): \K_1\to \cB_j
\]
is analytic and it has an analytic extension into whole complex
plane without zero.
Moreover, from \er{R01} we have the following estimate
$$
\|G(k)\|_{\cB_2}\le{\|\a\|\|\b\|\/|k|^2},\qqq k\in \ol\K_1\sm\{0\}.
$$
Moreover, we obtain
$\pa R_0(k)=\F_s^* \r_{ik}\r_k\F_s$, where $\r_k(\x)$
is the multiplication by
$(\x^2-k^2)^{-1}\x^{1\/2}$ and we have
\[
\lb{arb2}
\|G(k)\|_{\cB_1}\le
\big\|\a\F_s^* \r_{ik}\big\|_{\cB_2}
\big\|\r_{k}\F_s\b\big\|_{\cB_2}
\le
{2\|\a\|\|\b\|\/\pi}\|\r_{ik}\|
\|\r_{k}\|\le
{\|\a\|\|\b\|\/2\Re k\Im k},
\]
$k\in\K_1,$
since $\int_0^\iy \x|\x^2\pm k^2|^{-2}d\x\le{\pi\/4\Re k\Im k}$.

\subsection{Resolvent estimates}
\lb{ssY0Y}
The operator $H_0$ is self-adjoint  on  the form domain
$
\mD(H_0)=\{y,y''\in L^2(\R_+),y(0)=0\}.
$
The quadratic form $(Vy,y)$ is defined by
$$
(Vy,y)=-(2py',y')+(qy,y),\qq y\in \mD(H_0),
$$
 where
$(\cdot,\cdot)$ is the scalar product in $L^2(\R_+)$.
Then the standard arguments  give
\[
\lb{VE}
|(Vy,y)|\le{1\/2}\|y''\|^2+C\|y\|^2\qqq
\forall \qq y\in \mD(H_0),
\]
see e.g., \cite{K03},
for some constant $C>0$.
Then the KLMN theorem (see \cite[Th~X.17]{RS75}) yields
that there exists a unique self-adjoint operator  $H=H_0+V$
with the form domain, which coincide with $\mD(H_0)$, and
\[
\lb{qf4}
(Hy,y_1)=(H_0y,y_1)+(Vy,y_1)\qq\forall\qq
y,y_1\in \mD(H_0),
\]

In order to obtain resolvent estimates we need to discuss
the operator $Y_0$.
The definitions \er{defVj}, \er{Y0} imply
\[
\lb{exrepY0}
Y_0=
\ma (2p)^{1\/2}\pa R_0\pa |2p|^{1\/2}&(2p)^{1\/2}\pa R_0|q|^{1\/2}\\
q^{1\/2} R_0\pa |2p|^{1\/2}&q^{1\/2} R_0 |q|^{1\/2}\am,
\]
where $p^{1\/2}=|p|^{1\/2}\sign p$.
We introduce the operator-valued function $Y$ by
\[
\lb{defJJ0}
Y(k) =V_2R(k)V_1,\qqq k\in \K_1.
\]
This operator satisfies the standard equation:
\[
\lb{2.5}
(I-Y(k))(I+Y_0(k))=I\qqq \forall \ k\in \K_1\sm \s_d,
\]
where $\s_d$ is the set
of zeros of the function $D$ in $\ol\K_1$.

\begin{lemma}
\label{TY}
Let $p,q\in\cH_0$. Then

i) The operator $Y_0(k)\in \cB_j, j=1,2$ for each $k\in \K_1$,
the operator-valued
function $Y_0: \K_1\to \cB_j$ is analytic and has an analytic
extension into the whole complex plane without zero.
The operator-valued function $kY_0(k)$ is entire.
Moreover, $Y_0$ satisfies
\[
\lb{estY0op}
\|Y_0(k)\|_{\cB_2}\le{C\/|k|},\qq k\in\ol\K_1,
\]
\[
\lb{estY0}
\|Y_0(k)\|_{\cB_1}\le 2(2\|p\|_1+\|q\|_1)
\Big({1\/\Re k}+{1\/\Im k}\Big)\Big(1+{1\/|k|}\Big)^2,\qq
k\in\K_1
\]
for some constant $C=C(p,q)>0$.

ii) Each $Y(k)\in\cB_j,j=1,2,k\in\K_1\sm\s_d$ and  the operator-valued
function $Y: \K_1\sm\s_d\to \cB_j$ is analytic
and has a meromorphic  extension into the
whole complex plane.
Moreover, $Y$ satisfies
\[
\lb{asYk}
\|Y(k)\|_{\cB_2}\le{O(1)\/|k|},
\]
\[
\lb{estY2}
\|Y(k)-Y_0(k)\|_{\cB_2}\le {O(1)\/|k|^2},
\]
as $ k\in\ol\K_1, |k|\to\iy$.
\end{lemma}

{\bf Proof.}
i) The definition \er{Y0} and the identity \er{R01} yield \er{estY0op}.
Substituting the identities \er{R0r0}
into \er{exrepY0} and using the facts about the mappings $g,G$
in \er{ga}, \er{Ga}  we deduce that
 the operator-valued
function $Y_0: \K_1\to \cB_1$ is analytic and has an analytic
 extension into the whole complex plane without zero.
 The asymptotics \er{R0at0} shows that the
operator-valued function $kY_0(k)$ is entire.

Using  the estimates \er{arb1} we obtain for $\Im k>0$:
$$
\begin{aligned}
\|p^{1\/2}r_0(k)|p|^{1\/2}\|_{\cB_1}\le {4\|p\|_1\/\Im k},\qqq
\|q^{1\/2}r_0(k)|q|^{1\/2}\|_{\cB_1}\le {2\|q\|_1\/\Im k},\\
\|p^{1\/2}r_0(k)|q|^{1\/2}\|_{\cB_1}\le
{2\sqrt2(\|p\|_1\|q\|_1)^{1\/2}\/\Im k},
\end{aligned}
$$
 and the similar estimates with $r_0(ik)$ .
These estimates and the relations \er{arb2}, \er{exrepY0} give
$$
\|Y_0(k)\|_{\cB_1}\le (\|2p\|_1^{1\/2}+\|q\|_1^{1\/2})^2
\Big({1\/\Re k}+{1\/\Im k}\Big)\Big(1+{1\/|k|}\Big)^2,
$$
which yields \er{estY0}.

ii) For $k\in\K_1\sm \s_d$ identity \er{2.5} gives
\[
\lb{idYY0}
Y(k)=I-(I+Y_0(k))^{-1}=Y_0(k)(I+Y_0(k))^{-1}\in\cB_j,\qq j=1,2,
\]
and, since $Y_0(k)$ is analytic in $\K_1$, $Y(k)$ is analytic
in $\K_1\sm \s_d$.
Due to the analytic Fredholm theorem, see \cite[Th VI.14]{RS72},
the function $Y(k)$ has a meromorphic  extension into the
whole complex plane.
The estimate \er{estY0op} implies the asymptotics \er{asYk}.
Moreover,
$$
Y(k)-Y_0(k)=Y_0(k)\big((I+Y_0(k))^{-1}-I\big),
$$
which yields \er{estY2}.
\BBox

\section{ \lb{Sec3} The scattering matrix and the determinant}
\setcounter{equation}{0}

\subsection{The spectral representation for $H_0$}
The unitary transformation \er{Ft} carries over  $H_0$ into multiplication by
$k^4$ in $L^2(\R_+,dk)$:
$$
(\F_s  H_0\F_s ^* \wt f)(k) =k^4\wt f(k),\qq k>0,\qqq\where\qq \wt f(k)=(\F_sf)(k).
$$
Define a functional $\p_1(k) : L^2(\R_+)\oplus L^2(\R_+)\to \C$ by
\[
\lb{defPsi}
\p_1(k)f =(\F_s V_1 f)(k)=\sqrt{2\/\pi}\int_0^\iy
\Big(-k|2p(x)|^{1\/2}\cos kx,
|q(x)|^{1\/2}\sin kx\Big)f(x)dx,\qq k>0,
\]
$f\in L^2(\R_+)\oplus L^2(\R_+)$.
Define an operator $\p_2(k):\C\to L^2(\R_+)\oplus L^2(\R_+), k>0,$
by
\[
\lb{P*}
\p_2(k)c=V_2\sqrt{2\/\pi}\sin (kx) c=\sqrt{2\/\pi}\ma k(2p(x))^{1\/2}\cos kx \\
q(x)^{1\/2}\sin kx \am c,
\qq c\in\C.
\]
The operator-valued function $\p_j(k), k\in
\R_+, j=1,2$ have analytic extensions from $\R_+$ into the whole complex plane.
Then we can introduce the operators $\P_1(k)$ and $\P_2(k)$ by
\[
\lb{defPsi12}
 \P_1(k)=\ma\p_1(ik)\\ \p_1(k)\am,\qqq
 \P_2(k)=\Big(i\p_2(ik), \p_2(k)\Big),\qqq k\in\C.
\]

\begin{lemma}
\label{TPk} Let $p,q\in\cH_0$. Then  the operator-valued functions
$\p_j(k), j=1,2$, are entire and satisfy
\[
\label{Pe1}
\|\p_j(k)\|\le \sqrt{2\/\pi}
\Big(|k|^2\|2p\|_1+\|q\|_1\Big)^{1\/2}e^{\g |\Im k|}.
\]
\end{lemma}

 {\bf Proof.} The functional
$\p_1(k), k\in \C$ satisfies
$$
\begin{aligned}
\p_1(k)f=\sqrt{2\/\pi}\int_0^\g \Big(-k|2p(x)|^{1\/2}\cos kxf_1(x)
+|q(x)|^{1\/2}\sin kx f_2(x)\Big)dx,
\\
\qq |\p_1(k)f|^2\le {2e^{2\g|\Im k|}\/\pi} \Big(2|k|^2\|p\|_1\|f_1\|^2
+\|q\|_1\|f_2\|^2\Big),\qqq f=(f_1,f_2)\in L^2(\R_+)\oplus L^2(\R_+),
\end{aligned}
$$
which yields  \er{Pe1} for  $\p_1$.
Similarly,
$$
\|\p_2(k)c\|^2={2|c|^2\/\pi}
\int_0^\g \Big(|k|^2|2p(x)||\cos kx|^2
+|q(x)||\sin kx|^2\Big)dx,\qq c\in\C,
$$
which yields  \er{Pe1} for  $\p_2$.
\BBox

\medskip

For $k\in\C\sm\{0\}$ we
introduce finite rank operators
$P_1(k),P_2(k) $ on $L^2(\R_+)\os L^2(\R_+)$ by
\[
\lb{defP}
\begin{aligned}
P_1(k)=c_k\p_2(k)\p_1(k), \qqq c_k={\pi\/i2k^3},
\\
P_2(k)=P_1(ik)+P_1(k)=c_k \P_2(k)\P_1(k)
=c_k\big(i\p_2(ik)\p_1(ik)+\p_2(k)\p_1(k)\big).
\end{aligned}
\]
Below we need the following simple identities.

\begin{lemma}
Let $p,q\in\cH_0$ and let $k\in\C\sm\{0\}$. Then the operators
$P_1(k),P_2(k)$ satisfy
\[
\lb{cP1}
P_1(k)=Y_0(ik)-Y_0(k),
\]
\[
\lb{cP2}
P_2(k)= Y_0(-k)-Y_0(k).
\]
\end{lemma}

\no {\bf Proof.}
The identity \er{R0r0} implies
\[
\lb{R0kik}
R_0(k)-R_0(ik)={r_0(k)-r_0(-k)\/2k^2}.
\]
The identity \er{h1} shows that the kernel
of the integral operator $r_0(k)-r_0(-k)$
has the form
\[
\lb{r0kik}
r_0(x,y,k)-r_0(x,y,-k)=
{i\/k}\big(\cos k(x-y)-\cos k(x+y) \big)
={2i\/k}\sin kx\sin ky.
\]
The definitions \er{defPsi}, \er{P*} imply
$$
(\p_2(k)\p_1(k)f)(x)=
{2\/\pi}V_2\Big(\sin kx\int_0^\iy \sin ky (V_1f)(y)dy\Big).
$$
The identities \er{R0kik}, \er{r0kik} and the definition \er{Y0}
give
$$
Y_0(ik)-Y_0(k)=V_2(R_0(ik)-R_0(k))V_1
=c_k\p_2(k)\p_1(k),
$$
which yields the identity \er{cP1}.
The identities
$$
Y_0(-k)-Y_0(k)=Y_0(-k)-Y_0(ik)+Y_0(ik)-Y_0(k)=P_1(ik)+P_1(k)
$$
give \er{cP2}.
\BBox

\subsection{The scattering matrix.}
It is well known that the wave operators $W_\pm=W_\pm(H,H_0)$ for
the pair $H_0, H$, given by
$$
W_\pm=s-\lim e^{itH}e^{-itH_0} \qqq \as \qqq t\to \pm\iy,
$$
exist and  are complete, i.e., $\Ran W_\pm=\mH_{ac}(H)$.
The scattering operator $S=W_+^*W_-$ is unitary. The operators $H_0$
and $S$ commute and thus are simultaneously diagonalizable:
\[
\lb{DLH0}
L^2(\R_+)=\int_{\R_+}^\oplus \mH_\l d\l,\qqq
H_0=\int_{\R_+}^\oplus\l I_\l d\l,\qqq S=\int_{\R_+}^\oplus
S(\l^{1\/4})d\l;
\]
here $I_\l$ is the identity in the fiber space $\mH_\l=\C$ and
$S(k),k=\l^{1\/4}>0$
is the scattering matrix (which is a  scalar function
in our case) for the pair $H_0, H$.
The stationary representation for the scattering
matrix has the form
\[
\label{S}
S(k)=1+c_k\cA(k), \qqq k\in\R_+\sm\s_d,
\]
see e.g. \cite[Ch~XI.6]{RS79}, where $\s_d$ is the set
of zeros of the function $D$ in $\ol\K_1$,
and the "modified scattering amplitude" $\cA(k)$ is given by
\[
\lb{ScA}
\begin{aligned}
&\cA=\cA_0-\cA_1, \\
&\cA_0=\p_1\p_2,
\qqq
\cA_1 =\p_1Y\p_2,
\end{aligned}
\]
where $\p_1,\p_2$ are defined by \er{defPsi}, \er{P*}.

\begin{lemma}
\label{TA} Let $p,q\in\cH_0$. Then the function
$\cA_0(k)$ is continuous  in $\R_+$, it has an
analytic extension onto the whole complex plane and satisfies
\[
\label{A0i}
\cA_0(k)={1\/\pi}\Big(q_0-2p_0k^2
-\int_0^\iy \big(2k^2p(x)+q(x)\big)\cos 2kx dx\Big)\qqq
\forall\qq k\in\C,
\]
where
$
f_0=\int_0^\iy f(x)dx.
$
The function $\cA_1(k)$
is continuous  in $\R_+\sm\s_d$ and it has a meromorphic
extension onto the whole complex plane.
Moreover, the functions $\cA_0(k),\cA_1(k)$ satisfy
\[
\lb{ascA0kik}
\cA_0(k)=e^{2\g\Im k}O(k^2),\qqq
\cA_0(ik)=e^{2\g\Re k}O(k^2),
\]
\[
\lb{ascS11A}
\cA_1(k)=e^{2\g\Im k}O(k),
\]
as $|k|\to\iy, k\in\ol\K_1$.

\end{lemma}

\no {\bf Proof.} The operator-valued functions
$\p_1(k),\p_2(k)$ are continuous
in $\R_+$ and they have analytic
extensions onto the whole complex plane. Then
the function $\cA_0(k)$
is continuous  in $\R_+$ and it has an analytic
extension onto the whole complex plane.
The definitions  \er{defPsi}, \er{P*} and \er{ScA}
give
$$
\cA_0(k)
={2\/\pi}\int_0^\iy \big(-2k^2 p(x)\cos^2kx+q(x)\sin^2kx\big)dx,
$$
which yields the identity \er{A0i}.
This identity implies the asymptotics \er{ascA0kik}.

Due to Lemma~\ref{TY}~ii)
the function $\cA_1(k)$ is continuous
in $\R_+\sm\s_d$ and it has
a meromorphic extension onto the whole complex plane.
The definition \er{ScA} and the estimates \er{asYk} and \er{Pe1} give
the asymptotics \er{ascS11A}.
\BBox

\subsection{Properties of the determinant}
Lemma~\ref{TY}~i) shows that
$Y_0(k)\in\cB_1$, then the determinant
$D(k)=\det(I+Y_0(k)), k\in\K_1$ is well defined.

\begin{lemma}
\lb{TD1}
Let $p,q\in\cH_0$. Then

i) The determinant $D(k)=\det
(I+Y_0(k))$ is analytic in  $\K_1$ and  has  an analytic  extension
from $\K_1$ onto the whole complex plane
without zero, such that the function $kD(k)$ is entire.

ii)  The function $D(k)$ is real on the line $e^{i{\pi\/4}}\R$.
\end{lemma}

\no {\bf Proof.} i) Due to Lemma~\ref{TY}~i) the operator-valued
function $Y_0(k)$, and then the determinant $D(k)$, is analytic in
$k\in \K_1$ and has  an analytic extension from $k\in \K_1$ onto the
whole complex plane without zero. It is proved in \cite{BK16} that
the function $kD(k)$ is entire.

ii) The identity \er{R01} shows that $R_0(k)$
is real on the line
$e^{i{\pi\/4}}\R$, then $Y_0(k)$ is real also.
Therefore, $D(k)$ is real on this line.
\BBox

\medskip

The estimates \er{estY0} give $\|Y_0(k)\|_{\cB_1}=O(k^{-1})$ as
$k\to e^{i{\pi\/4}}\infty$.
We can define the branch $\log D(k)$ for $k\in\K_1$
and $|k|$ large enough, by
$$
\log D(k) = o(1)\qqq
\as\qqq k\to e^{i{\pi\/4}}\infty.
$$
We need the following standard results.

\begin{lemma}
\lb{TD2}
Let $p,q\in\cH_0$.
Then the following identity holds true:
\[
\lb{TrY0}
\Tr Y_0(k)
=-{(1+i)p_0-i\wh p(k)-\wh p(ik)\/2k}-{(1-i)q_0+i\wh q(k)-\wh q(ik)\/4k^3},
\]
for any $k\in \K_1$,
where
$
\wh f(k)=\int_0^\iy e^{2ikx}f(x)dx.
$
Moreover, the function $\log D(k)$ satisfies
\[
\lb{2.13}
 |\log D(k)+\sum _{n=1}^{N}{1\/n}\Tr (-Y_0(k))^n|\le
{C_1\/|k|^{N+1} },\qqq \forall  \qq N\ge 1,
\]
\[
\lb{2.12}
\log D(k)=-\sum _{n=1}^\iy {1\/n}\Tr (-Y_0(k))^n, \ \ \
\]
for any $k\in \K_1,|k|$ large enough, and for some constant $C_1>0$,
where the series converges absolutely and uniformly on
any compact subset of $\K_1$.
Furthermore, the function $\log D$ satisfies the asymptotics
 \[
\lb{aD1}
 \log D(k)=-\fr{(1+i)p_0+o(1)}{2k}
\qqq \as \qq |k|\to \iy,\qq k\in \ol\K_1
\]
uniformly in $\arg k\in[0,{\pi\/2}]$.
\end{lemma}

\no {\bf Proof.} Let $k\in\K_1$.
The identities \er{R0r0}, \er{exrepY0} imply
$$
\Tr Y_0(k)
=\int_0^\iy\Big(-\big(r_0(x,x,k)+r_0(x,x,ik)\big)p(x)
+{(r_0(x,x,k)-r_0(x,x,ik))q(x)\/2k^2}\Big)dx.
$$
Then the identity \er{h1} gives \er{TrY0}.
The estimate \er{estY0op} gives
\[
\lb{estY0n}
\big|\Tr(Y_0(k))^n\big|\le\|Y_0(k)\|_{\cB_2}^{n}
\le\Big({C\/|k|}\Big)^{n},\qqq n\ge 2
\]
for some constant $C>0$.
Then the series \er{2.12} converges absolutely
and uniformly and
it is well-known that
the sum is equal to $\log D(k)$
(see \cite[Lm XIII.17.6]{RS78}).
The estimates \er{estY0n} imply \er{2.13}.
The relations \er{TrY0}, \er{2.13} give the asymptotics \er{aD1}.
\BBox

\subsection{Identities for the determinant and S-matrix}
We will determine asymptotics of the determinant in the complex plane. In
the case of the Scr\"odinger operator it is sufficiently to obtain
the asymptotics of the determinant $d(k)$ and the scattering matrix
$s(k)$ in $\C_+$. Then using the identity \er{cS(k)2o} we obtain the
asymptotics in $\C_-$. In the case of fourth order operators the
similar arguments give the asymptotics of the determinant in the
domains $\K_1,\K_2$ (and, by the symmetry, in $\K_4$). In order to
obtain the asymptotics in the domain $\K_3$ we need some additional
analysis, more complicated than for $\K_1,\K_2$.
The corresponding analysis for third order operators
was carried out in \cite{K16}.

Introduce the $2\ts 2$ matrix-valued function $\O(k),k\in\K_1$ by
\[
\lb{dcB1}
\O=1+c_k(\O_0-\O_1),\qqq \O_0=\P_1\P_2,\qq \O_1=\P_1Y\P_2,
\]
where $\P_1,\P_2$ are defined by \er{defPsi12}.
The function $\O_0$ has an analytic extension
from $\K_1$ into the whole complex plane
and the functions $\O_1$ and $\O$
have meromorphic extensions from $\K_1$ into the
whole complex plane.

The identity \er{cS(k)} below is similar to the relation \er{cS(k)2o}
for Schr\"odinger operator. It gives an exact formula
for an analytic extension of the determinant $D$ in
the domain $\K_2$.
In order to get the analytic extension in the domain $\K_3$
we use the identity \er{cT(k)}. It is a crucial point for our
consideration.

\begin{lemma} Let $p,q\in\cH_0$, and let $k\in\C\sm\{0\}$. Then
the determinant $D$ satisfies
\[
\lb{cS(k)}
D(ik)=D(k)S(k),
\]
\[
\lb{cT(k)}
D(-k)=D(k)\det\O(k),
\]
where  the function $\O$ is given by the definition \er{dcB1},
and the $S$-matrix $S(k)$ is continuous on $\R_+$.

\end{lemma}

\no {\bf Proof.}
The identities \er{cP1}  and \er{2.5} give
$$
D(ik)=\det\big(1+Y_0(ik)\big)=\det\big(1+Y_0(k)+P_1(k)\big)
=D(k)\det\Big(1+\big(1-Y(k)\big)P_1(k)\Big).
$$
The definition \er{defP}
yields
$$
D(ik)=D(k)\det\Big(1+c_k\big(1-Y(k)\big)\p_2(k)\p_1(k)\Big).
$$
Then the definitions \er{S}, \er{ScA}
and the identity \er{2.2} imply the identity \er{cS(k)}.

Similarly, the identities \er{cP2}  and \er{2.5}  give
$$
D(-k)=\det\big(1+Y_0(-k)\big)
=\det\big(1+Y_0(k)+P_2(k)\big)
=D(k)\det\Big(1+\big(1-Y(k)\big)P_2(k)\Big).
$$
The definition \er{defP}
yields
$$
D(-k)=D(k)\det\Big(1+c_k\big(1-Y(k)\big) \P_2(k)\P_1(k)\Big).
$$
Then the identity \er{2.2} implies the identity \er{cT(k)}.

Due to Lemma~\ref{TA}, the $S$-matrix $S(k)$ is continuous in
$k\in \R_+\sm\s_d$ and it has a
meromorphic extension from $\R_+$ onto $\C$.
Moreover, if $k\in\s_d\cap\R_+$,
then $k$ is a zero of the functions $D(ik)$
and $D(k)$ of the same multiplicity. Due to the identity
\er{cS(k)}, $S(k)$ is continuous
at the point $k\in\s_d$.
Therefore, $S(k)$ is continuous on $\R_+$.
\BBox

\begin{lemma}
Let $p,q\in\cH_0$.
Then the function $\O_0=\P_1\P_2$ satisfies the identity
\[
\lb{idcB01}
\O_0(k)=\ma i\cA_0(ik)& \mB(k)\\
i\mB(k)&\cA_0(k)\am,\qqq k\in\C,
\]
where $\cA_0$ is defined by \er{ScA} and
\[
\lb{defmB}
\mB(k)={2\/\pi}\int_0^\iy
\Big(-i2k^2p(x)\ch kx\cos kx
+iq(x)\sh kx\sin kx\Big) dx.
\]
The function $\mB$ satisfies
\[
\lb{asmBK1}
\mB(k)=e^{\g(\Re k+\Im k)}O(k^2)
\]
as $|k|\to\iy,k\in\ol\K_1$.
\end{lemma}

\no {\bf Proof.}
We prove the identity \er{idcB01}.
The definition \er{defPsi12} implies
\[
\lb{idOm0pr}
\O_0(k)=\P_1(k)\P_2(k)=\ma i\p_1(ik)\p_2(ik)& \p_1(ik)\p_2(k)\\
i\p_1(k)\p_2(ik)&\p_1(k)\p_2(k)\am.
\]
The definitions  \er{defPsi}, \er{P*} and \er{defmB} give
$$
\p_1(ik)\p_2(k)={2\/\pi}\int_0^\iy
\Big(-i2k^2p(x)\ch kx\cos kx
+iq(x)\sh kx\sin kx\Big) dx=\mB(k).
$$
and similarly $\p_1(k)\p_2(ik)=\mB(k)$.
Substituting these identities and the definition
\er{ScA} into \er{idOm0pr} we obtain
the identity \er{idcB01}.
The definition \er{defmB} yields the asymptotics \er{asmBK1}.
\BBox

\medskip

\section{ \lb{Sec4} Proof of the main Theorems and trace formulas in terms of
resonances}
\setcounter{equation}{0}

\subsection{Asymptotics of the determinant}
We prove our preliminary  Proposition \ref{T1} which gives
some properties of the determinant and
its asymptotics in the domain $\K_1$.

\medskip

\no {\bf Proof of Proposition \ref{T1}.}
i) The statement is proved in Lemma~\ref{TY}~i).

ii) Due to Lemma~\ref{TD1}, the function $D$ has an
analytic extension from $\K_1$ onto
$\C\sm\{0\}$,
it is real on the line
$e^{i{\pi\/4}}\R$ and the function $kD(k)$ is entire.
The asymptotics \er{aD1} yields  \er{asDK+}.
The function $D(k)$ has
a finite number of zeros in $\ol\K_1$, then
the operator $H$ has a finite number of eigenvalues.
\BBox

\medskip

The asymptotics of $D(k)$ in the domain $\K_1$
is known due to \er{asDK+}.
We analyze the function $D(k)$ in the domains $\K_2,\K_3$
by the following way.
We obtain the asymptotics
of $S(k)$ and $\O(k)$ in $\K_1$. Then
we use the identities \er{cS(k)}, \er{cT(k)}
in order to determine the asymptotics of $D(ik),D(-k)$
in $\K_1$, which gives the asymptotics of $D(k)$
in $\K_2,\K_3$.

\begin{lemma} Let $p,q\in\cH_0$ and let $k\in\ol\K_1,|k|\to\iy$. Then
\[
\lb{asSf}
S(k)=1+e^{2\g\Im k}O(k^{-1}),
\]
\[
\lb{estDik}
D(ik)=1+e^{2\g\Im k}O(k^{-1}),
\]
\[
\lb{estD-k}
D(-k)=e^{2\g\Im k}O(k^{-1})+e^{2\g\Re k}O(k^{-1})+
e^{2\g(\Re k+\Im k)}O(k^{-2})
\]
uniformly in $\arg k\in[0,{\pi\/2}]$.
\end{lemma}

\no {\bf Proof.}
Let $|k|\to\iy, k\in\ol\K_1$.
Substituting the asymptotics \er{ascA0kik} and
\er{ascS11A} into the definition \er{ScA} we obtain
$\cA(k)=e^{2\g\Im k}O(k^2)$. Then
the identity \er{S} gives the asymptotics \er{asSf}.
Substituting the asymptotics \er{asDK+}, \er{asSf}  into
\er{cS(k)} we obtain
the asymptotics  \er{estDik}.

Substituting the asymptotics \er{ascA0kik} and \er{asmBK1}
into the identity \er{idcB01} we obtain
\[
\lb{asOm0}
\O_0(k)=\ma e^{2\g\Re k}O(k^2)&e^{\g(\Im k+\Re k)}O(k^2)\\
e^{\g(\Im k+\Re k)}O(k^2)&e^{2\g\Im k}O(k^2)\am.
\]
The definitions \er{defPsi12}, \er{dcB1} give
$$
\O_1(k)=\ma i\p_1(ik)Y(k)\p_2(ik)& \p_1(ik)Y(k)\p_2(k)\\
i\p_1(k)Y(k)\p_2(ik)&\p_1(k)Y(k)\p_2(k)\am
$$
Then the estimates \er{asYk} and \er{Pe1} imply
\[
\lb{ascS11}
\O_1(k)=\ma e^{2\g\Re k}O(k)&e^{\g(\Im k+\Re k)}O(k)\\
e^{\g(\Im k+\Re k)}O(k)&e^{2\g\Im k}O(k)\am.
\]
Substituting the asymptotics \er{asOm0} and
\er{ascS11} into the definition \er{dcB1}
we obtain
$$
\O(k)=\ma 1+e^{2\g\Re k}O(k^{-1})&e^{\g(\Im k+\Re k)}O(k^{-1})\\
e^{\g(\Im k+\Re k)}O(k^{-1})&1+e^{2\g\Im k}O(k^{-1})\am,
$$
which yields
\[
\lb{asSf2}
\det\O(k)=e^{2\g\Im k}O(k^{-1})+e^{2\g\Re k}O(k^{-1})+
e^{2\g(\Re k+\Im k)}O(k^{-2}).
\]
Substituting the asymptotics \er{asDK+}, \er{asSf2}
into the identity \er{cT(k)} we obtain the asymptotics \er{estD-k}.
\BBox

\subsection{Asymptotics of the counting function}
Recall that the function $D(k)$ is analytic in $\C\sm\{0\}$
and may have a simple pole at the point $k=0$.
We prove our main results.

\medskip

\no {\bf Proof of Theorem~\ref{CorEstN}}.
The asymptotics \er{asDK+}, \er{estDik}, \er{estD-k}
give the estimate \er{unifestD} in $\K_1,\K_2,\K_3$.
The symmetry
$D(k)=\ol{D(i\bar k)}$ imply the estimate
\er{unifestD} in $\K_4$.

The asymptotics \er{estDik}
gives
$D(k)=1+e^{-2\g\Re k}O(k^{-1})$
as $|k|\to\iy,k\in\K_2$. This yields
$|k(D(k)-1)|\le Ce^{-2\g\Re k}$ for all $k\in\K_2$
with modulus large enough.
Let $k_*\in\K_2$ be a resonance.
Then the identity $D(k_*)=0$ gives the estimate \er{estFD}.

We prove the estimate \er{estnr}.
Recall that the function $D(k)$ is analytic in $\C\sm\{0\}$
and may have a simple pole at the point $k=0$.
Let the function $F(k)=k^mD(k),m\le 1$, be entire and
satisfy $F(0)\ne 0$.
Let $\cN_F(r)$ be the number of zeros
of the function $F$ in the disc $|k|<r$ counted with multiplicity.
If $D(0)\ne 0$, then $\cN=\cN_F$,
if $k=0$ is a zero of $D$
of multiplicity $\ell$, then $\cN=\cN_F+\ell$.
It is sufficiently to prove the estimate \er{estnr}
for $\cN_F$.
The estimate \er{unifestD} gives
\[
\lb{esF}
\log|F(k)|\le\g \big((\Re k)_-+(\Im k)_-\big)+
C\log |k|
\]
for all $k\in\C,|k|$ large enough and
for some $C>0$.
Substituting the estimate \er{esF} into Jensen's formula
\[
\lb{JFg}
\int_0^r\fr{\cN_F(t)}{t}dt=
\fr{1}{2\pi}\int_0^{2\pi}\log |F(re^{i\theta})|d\theta
-\log|F(0)|,
\]
we obtain
$$
\int_0^r\fr{\cN_F(t)}{t}dt\le
-\fr{\g r}{\pi}\Big(
\int_{\fr{\pi}{2}}^{\pi}\cos\theta d\theta
+\int_{\pi}^{\fr{3\pi}{2}}
(\cos\theta+\sin\theta ) d\theta
+\int_{\fr{3\pi}{2}}^{2\pi}\sin\theta d\theta\Big)
+ C\log r=
\fr{4\g r}{\pi}+ C\log r
$$
for all $r>0$ large enough.
Then there exists
$$
\lim_{r\to+\iy} {1\/r}\int_0^r\fr{\cN_F(t)}{t}dt\le{4\g\/\pi}.
$$
The estimate \er{estnr} follows from the following
well known result,
see, e.g., \cite[Lm II.4.3]{L71}:

{\it
Let $\cN_F(t)$ be non-decreasing function on $\R_+$,
$\cN_F(t)=0$ as $0\le t<\ve$ for some $\ve>0$, and let the function
$$
I(r)=\fr{1}{r}\int_0^r\fr{\cN_F(t)}{t}dt,\qqq r\in\R_+,
$$
has the limit as $r\to\iy$.
Then
$
\cN_F(r)=r(I(r)+o(1)).
$
}
\BBox

\subsection{Scattering phase and trace formulas}
Now we discuss the Hadamard factorization for the Fredholm determinant
$D$.
The function $kD(k)$ is entire, then
$$
D(k)={\a\/k^{m}}\big(1+\b k+O(k^2)\big)\qqq \as \qq |k|\to 0,\qqq
m\le 1,
$$
for some $\a,\b\in\C$.
Let $\z_n,n\in\N$, be the zeros of the function $D$ in $\C\sm\{0\}$
labeled by $0<|\z_1|\le|\z_2|\le...$  counting with multiplicities.
The estimate \er{unifestD} provides
the standard Hadamard factorization
\[
\lb{HadF}
D(k)={\a\/k^{m}} e^{\b k}
\lim_{r\to\iy}
\prod_{|\z_n|<r}\Big(1-{k\/\z_n}\Big)e^{k\/\z_n},
\]
absolutely and uniformly on any compact
subset in $\C\sm\{0\}$.

\medskip

\no {\bf Remark.} It is proved in \cite{BK16}
that $\b=(i-1)\g$ in the case $(p,q)\in\cH_1\ts\cH_0$, $p(\g-0)\ne 0$.

\medskip

The S-matrix $S(k), k\in\R_+$ is a complex number and
$|S(k)|=1$. Thus we have
\[
 \label{Sx}
 S(k)=e^{-2\pi i\f_{sc}(k)},\qqq k\in\R_+,
\]
where $\f_{sc}(k)$ is a scattering phase.
The function $S(k)$ is continuous on $\R_+$
and the asymptotics \er{asSf} shows that $S(k)=1+O(k^{-1})$ as $k\to +\iy$.
If we assume that the function $\f_{sc}(k)$ is also continuous on $\R_+$,
then formula \er{Sx} uniquely determines
$\f_{sc}(k),k>0,$ by $\f_{sc}(k)={i\/2\pi }\log S(k)$
and the
asymptotics $\f_{sc}(k)=O(k^{-1})$ as $k\to +\iy$.

Our next results concern the trace formula in terms of resonances.
Trace formulas for one-dimensional  Schr\"odinger operators
in terms of resonances were determined in \cite{K04}
and for third order operators in \cite{K16}.
Here we use the approach from \cite{K04}.

\begin{theorem}
\lb{ThRes}
Let
$(p,q)\in\cH_0$.
Then  the following trace formulas hold true:
\[
\lb{expF'}
4k^4\Tr (R_0(k)-R(k))=-m+\b k+k^2\lim_{r\to\iy}\sum_{|\z_n|<r}
{1\/\z_n (k-\z_n)},\qqq k\in\K_1\sm\s_d,
\]
the series converges absolutely and uniformly on any compact subset in
$\K_1\sm\s_d$,
\[
\lb{expphi'}
\phi_{sc}'(k)
={1\/2\pi i}\lt((1-i)\b
+\sum_{n=1}^\iy{k\/\z_n}\Big({1\/ik-\z_n}
+{1\/k-\z_n}\Big)\rt),\qqq k\in\R_+\sm\s_d,
\]
the series converges absolutely and uniformly on any compact subset in
$\R_+\sm\s_d$.
\end{theorem}

\no {\bf Proof.} Let $k\in\K_1\sm\s_d$.
The definitions \er{defVj}, \er{R01} show that the operators
$VR_0(k),V_2R_0(k)$ and $R_0(k)V_1$
are Hilbert-Schmidt. Then the operator
$$
R_0(k)-R(k)=R_0(k)VR(k)=R_0(k)VR_0(k)-R_0(k)VR_0(k)VR(k)
$$
is trace class.
Due to the identities \er{2.2}, \er{2.3}, \er{2.5} and
$Y_0'(k)=4k^3V_2R_0^2(k)V_1$,
the derivative of $D$ satisfies
\[
\lb{idTrRR0}
{1\/4k^3}{D'(k)\/D(k)}=\Tr \big((I+Y_0(k))^{-1}V_2R_0^2(k)V_1\big)=
\Tr R_0(k)VR(k)=\Tr (R_0(k)-R(k)).
\]
The identity \er{HadF} gives
\[
\lb{expF'1}
{D'(k)\/D(k)}=\b-{m\/k}+k\lim_{r\to\iy}\sum_{|\z_n|<r}
{1\/\z_n (k-\z_n)}.
\]
This identity together with \er{idTrRR0} yields
the trace formula \er{expF'}.

The function $S(k)$ is continuous in $\R_+$,
has a meromorphic extension onto the whole complex plane
and, due to equations \er{Sx} and \er{cS(k)},
it satisfies the identities
$$
e^{-2\pi i\phi_{sc}(k)}=S(k)={D(ik)\/D(k)},
\qqq\forall\qq k>0.
$$
Differentiating  this identity
we obtain
$$
-2\pi i\phi_{sc}'(k)
=i{D'(ik)\/D(ik)}-{D'(k)\/D(k)}.
$$
Then equation \er{expF'1} implies \er{expphi'}.
\BBox

\section{\lb{Sec5} Euler-Bernoulli operators and proof of Theorem~\ref{ThEB}}
\setcounter{equation}{0} \lb{sect55}

\subsection{The Liouville type transformation}
We consider the Euler-Bernoulli operator
\[
\lb{defcE}
\cE u={1\/b}(au'')'', \qqq
\]
acting on $L^2(\R_+,b(x)dx)$, with the boundary conditions
\[
\lb{ebdc}
u=0,\qqq \text{and} \qq u''+{a'\/5a}u'=0\qq
\text{at}\qq x=0,
\]
where the  coefficients $a,b$ are positive, $a=b=1$ outside the unit
interval and satisfy
\[
\lb{abc}
a-1,b-1\in \cH_4,\qq
\Big({3a'\/a}+{5b'\/b}\Big)\Big|_{x=0}=0.
\]

Now we consider the Liouville type transformation of the operator
$\cE$ into the operator $H$, defined by \er{a.1}, \er{4g.dc}
with specific $p,q$ depending on $a,b$. In
order to define this transformation we introduce the new variable
$t\in\R_+$ by
\[
\lb{tx}
t=t(x)=\int_{0}^x\Big({b(s)\/a(s)}\Big)^{1\/4}ds,\qqq\forall\qq x\in\R_+.
\]
Let $x=x(t)$ be the inverse
function for $t(x), x\in\R_+$. Introduce the unitary
transformation
$U:L^2(\R_+,b(x)dx) \to L^2(\R_+,dt)$ by
\[
\lb{defU}
u(x)\mapsto (Uu)(t)=a^{1\/8}(x(t))b^{3\/8}(x(t))u(x(t)).
\]

Introduce the functions $\a(t),\b(t),t\in\R_+$, by
\[
\lb{abxe}
\a(t)={1\/a(x(t))}{da(x(t))\/dt},\qqq \b(t)={1\/b(x(t))}{db(x(t))\/dt}.
\]
Then the functions $\a,\b\in L^1(\R_+)$ are real, compactly supported
and satisfy
$$
\stackrel{\ldots}{a},\stackrel{\ldots}{\b}\in L^1(\R_+),\qqq
\where\qq \dot f={df\/dt}.
$$

Let the
operator $H$ be defined by \er{a.1},
where the coefficients $p(t),q(t),t\in\R_+$, have the forms
\[
\lb{peb}
p=-{\dot\e_0+\vk\/2},
\]
\[
\lb{kapx}
\vk={5\a^2+5\b^2+6\a\b\/32},\qqq \e_0={3\a+5\b\/4},
\]
and
\[
\lb{qeb}
q={d\/dt}\big((\dot \e_2+\e_2^2)\e_1-\ddot \e_1\big)
+\big((\dot \e_2+\e_2^2)\e_1-\ddot \e_1\big)\e_1,
\]
$$
\e_1={\a+3\b\/8},\qqq \e_2={3\a+\b\/8}.
$$
Note that the coefficients
$(p,q)\in \cH_2\ts\cH_0$, where $\g$
is given by the definition \er{gammaEB}.
The definition \er{kapx} shows that
\[
\lb{estvk}
\vk\ge {\a^2+\b^2\/16}\ge 0,
\]
moreover, $\vk=0$ iff $\a=\b=0$.
The proof of Theorem \ref{ThEB} is based on
this observation.

Let the coefficients $a,b$ be positive and satisfy the conditions
\er{abc}. Let the operator $\cE$ be defined by \er{defcE}
and let the operator $H$ be defined by \er{a.1},
where the coefficients
$p,q$ have the forms \er{peb}, \er{qeb}.
Repeating the arguments from \cite{BK15} we obtain that
the operators $\cE$ and $H$ are unitarily equivalent and
satisfy:
\[
\lb{eqEH}
\cE=U^{-1}HU.
\]
where the operator $U$ is defined by \er{defU}.

\medskip

\no {\bf Proof of Corollary \ref{CorEB}.}
The definitions \er{tx}, \er{peb}, \er{qeb} show that
$(p,q)\in \cH_2\ts\cH_0$, where $\g$
is given by \er{gammaEB}.
This yields the statement.
\BBox

\medskip

The following Lemma is a corollary of Proposition~\ref{T1}.
Here we determine asymptotics of the determinant in the domain
$\K_1$, which is crucial for the proof of Theorem \ref{ThEB}.

\begin{corollary}
Let the coefficients $a,b$ be positive and satisfy the conditions
\er{abc}.
Then the determinant $D(k)$ satisfies
\[
\lb{asdetGEB}
D(k)=1+\fr{1+i}{8k}\int_{0}^\iy\vk(t) dt+{O(1)\/k^2}\qqq
\as \qqq |k| \to \infty, \qq  k\in\ol\K_1
\]
uniformly in $\arg k\in[0,{\pi\/2}]$, where $\vk(t)$
is given by the definition \er{kapx}.
\end{corollary}

\no {\bf Proof.}
Due to the last condition in \er{abc1} we have $\e_0(0)=0$.
Identity \er{peb} gives
$$
p_0=\int_0^\iy p(t)dt=-{1\/2}\int_0^\iy\vk(t) dt.
$$
Substituting this identities into the asymptotics
\er{asDK+} we obtain the asymptotics \er{asdetGEB}.
\BBox

\medskip

\no {\bf Proof of Theorem \ref{ThEB}.}
The proof uses the arguments of
Isozaki and Korotyaev \cite{IK12}.
Assume that the operator $\cE$ does not have any eigenvalues and
resonances. The identity \er{HadF} shows that
$D(k)=k^{-m}\a e^{\b k}$ in this case,
where $\a,\b\in\C$ and $m=0$ or $1$.
The asymptotics
\er{asDK+} implies
$D=1$. Then the second term in the asymptotics
\er{asdetGEB} vanishes, which yields $\int_0^\iy\vk(t) dt=0$.
Due to the estimate \er{estvk},
we have $\vk=0$ and $\a=\b=0$
in this case, then  $a=b=1$ on $\R_+$.

Conversely,
let $a=b=1$ on $\R_+$. Then the operator $\cE$, given by \er{defcE}, has the form
$\cE=\pa^4$. The definitions \er{Y0}, \er{a.2} show that $D=1$ in this case.
Then, due to the identity \er{HadF},
there are not any eigenvalues and resonances.
\BBox

\section{\lb{Sec6} Resonances for coefficients with jump
discontinuity and proof of  Theorem~\ref{ThAsCF}}
\setcounter{equation}{0}

\subsection{Asymptotics of auxiliary functions}
The function $D(k)$ has a finite number of zeros in the domain
$\K_1$. The identity \er{cS(k)} shows that $ik$
with large $|k|$ is a resonance in $\K_2$ iff
$k$ is a zero of the function $S(k)$ in $\K_1$.
Thus in order to determine asymptotics of resonances
in $\K_2$ we need to improve asymptotics of the scattering
matrix $S(k)$ in $\K_1$.
Similarly, the identity \er{cT(k)} shows that $-k$
with large $|k|$ is a resonance in $\K_3$ iff
$k$ is a zero of the function $\det\O(k)$ in $\K_1$.
Then in order to determine asymptotics of resonances
in $\K_3$ we have to improve asymptotics of the
function $\det\O(k)$ in $\K_1$. Moreover,
due to the symmetry of the determinant it is sufficiently
to consider in this case the domain
$$
\K_1^+=\big\{k\in\C:\arg k\in\big(0,\tf{\pi}{4}\big)\big\}.
$$

In the following Lemma we improve the asymptotics of
the functions $\cA_0$ and $\mB$, given by \er{ScA} and \er{defmB}
respectively.

\begin{lemma}
Let $(p,q)\in\cH_1\ts\cH_0$. Then the functions
$\cA_0$ and $\mB$ satisfy
\[
\lb{ascA0imp}
\cA_0(k)=-{ike^{-i 2k\g}\/2\pi}\big(p_++o(1)\big)+O(k^2),
\]
\[
\lb{ascA0impi}
\cA_0(ik)={ke^{2k\g}\/2\pi}\big(p_++o(1)\big)+O(k^2),
\]
as $|k|\to\iy, k\in\ol\K_1$ uniformly in $\arg k\in[0,{\pi\/2}]$,
\[
\lb{asmBimp}
\mB(k)=-{ik\/2\pi}(1+i)e^{(1-i) k\g}\big(p_++e^{-2 \g\Im k}O(1)+o(1)\big)+O(k^2),
\]
as $|k|\to\iy, k\in\ol\K_1^+$ uniformly in $\arg k\in[0,{\pi\/4}]$.
\end{lemma}

\no {\bf Proof.}
Let $|k|\to\iy, k\in\ol\K_1$.
The identity \er{A0i} implies
$$
\cA_0(k)=-{k^2\/\pi}\int_0^\iy p(x)e^{-i2kx} dx+O(k^2)
+e^{2\g\Im k}O(1).
$$
Integrating by parts we obtain \er{ascA0imp}. Similarly,
$$
\cA_0(ik)={k^2\/\pi}\int_0^\iy p(x)e^{2kx} dx+O(k^2)
+e^{2\g\Re k}O(1),
$$
which implies \er{ascA0impi}.

Let $k\in\ol\K_1^+,|k|\to\iy$.
The definition  \er{defmB} gives
$$
\mB(k)=-{ik^2\/\pi}\int_0^\iy p(x)
(e^{(1-i)kx}+e^{(1+i)kx})dx+O(k^2)+e^{\g(\Im k+\Re k)}O(1).
$$
Integrating by parts we obtain the asymptotics \er{asmBimp}.
\BBox

\medskip

Now we improve the asymptotics of
the functions $\cA_1$ and $\O_1$, given by \er{ScA} and \er{dcB1}
respectively.

\begin{lemma}
Let $(p,q)\in\cH_1\ts\cH_0$. Then the functions $\cA_1$ and $\O_1$
satisfy
\[
\lb{asA1im}
\cA_1(k)=e^{2\g\Im k}O(1),
\]
\[
\lb{asB1im}
\O_1(k)=\ma e^{2\g\Re k}O(1)&e^{\g(\Im k+\Re k)}O(1)\\
e^{\g(\Im k+\Re k)}O(1)&e^{2\g\Im k}O(1)\am
\]
as $|k|\to\iy, k\in\ol\K_1$ uniformly in $\arg k\in[0,{\pi\/2}]$.
\end{lemma}

\no {\bf Proof.}
The definitions \er{defPsi12}, \er{ScA}, \er{dcB1}
and the estimates
\er{estY2}, \er{Pe1} give
\[
\lb{asA1B1}
\begin{aligned}
\cA_1(k)=\p_1(k)Y_0(k)\p_2(k)+e^{2\g\Im k}O(1),
\\
\O_1(k)= \P_1(k)
Y_0(k)\P_2(k)
+\ma e^{2\g\Re k}O(1)&e^{\g(\Im k+\Re k)}O(1)\\
e^{\g(\Im k+\Re k)}O(1)&e^{2\g\Im k}O(1)\am ,
\end{aligned}
\]
as $|k|\to\iy, k\in\ol\K_1$.

Let $k\in\ol\K_1$.
The definitions \er{defPsi}, \er{P*} and
the identity \er{exrepY0} imply
$$
\p_1(k)Y_0(k)\p_2(k)=k^2a_1(k)-k\big(a_2(k)+a_3(k)\big)+a_4(k),
$$
where
$$
a_1=4\int_0^\iy p(x)\cos kx \Big(\int_0^\iy
{\pa R_0(x,y,k)\/\pa x}\big(p(y)\cos ky\big)'dy\Big)dx,
$$
$$
a_2=2\int_0^\iy q(x)\sin kx \Big(\int_0^\iy
R_0(x,y,k)\big(p(y)\cos ky\big)'dy\Big) dx,
$$
$$
a_3=2\int_0^\iy p(x)\cos kx\Big(\int_0^\iy
{\pa R_0(x,y,k)\/\pa x}q(y)\sin kydy\Big)dx,
$$
$$
a_4=\int_0^\iy q(x)\sin kx\Big(\int_0^\iy
R_0(x,y,k)q(y)\sin kydy\Big)dx.
$$
Using the identity $a_3=-a_2$ we obtain
\[
\lb{idcP0}
\p_1(k)Y_0(k)\p_2(k)=k^2a_1(k)+a_4(k).
\]
Let $|k|\to\iy, k\in\ol\K_1$.
The identity \er{R01} gives
\[
\lb{asA234}
a_4(k)={e^{2\g\Im k}O(1)\/k^3}.
\]
Moreover, the identity \er{R0r0} yields
\[
\lb{idA1}
a_1(k)=b_1(k)+b_2(k),
\]
where
$$
\begin{aligned}
b_1(k)=-2\int_0^\iy  p(x)\cos kx\int_0^\iy
r_0(x,y,k)p(y)\cos kydydx,
\\
b_2(k)=-2\int_0^\iy  p(x)\cos kx\int_0^\iy
r_0(x,y,ik)p(y)\cos kydydx.
\end{aligned}
$$
The identity \er{h1} and the integration by parts give
$$
b_1(k)
=-{i\/k}\int_0^\iy p(x)\cos kx
\Big(\int_0^\iy \big(e^{ik|x-y|}-e^{ik(x+y)} \big)p(y)\cos kydy\Big)dx
={e^{2\g\Im k}O(1)\/k^{2}},
$$
and similarly,
$$
b_2(k)={e^{2\g\Im k}O(1)\/k^{2}}.
$$
Then the identity \er{idA1} yields
\[
\lb{asA1}
a_1(k)={e^{2\g\Im k}O(1)\/k^2}.
\]
Substituting the asymptotics \er{asA234},
\er{asA1} into the identity \er{idcP0} we obtain
$$
\p_1(k)Y_0(k)\p_2(k)=e^{2\g\Im k}O(1).
$$
The asymptotics \er{asA1B1}  gives
the asymptotics \er{asA1im}.

The similar arguments show that
$$
\begin{aligned}
\p_1(ik)Y_0(k)\p_2(ik) =e^{2\g\Re k}O(1),
\\
 \p_1(ik)Y_0(k)\p_2(k) =e^{\g(\Im k+\Re k)}O(1),
\\
\p_1(k)Y_0(k)\p_2(ik) =e^{\g(\Im k+\Re k)}O(1).
\end{aligned}
$$
Substituting these asymptotics into the definition \er{defPsi12}
we obtain
$$
\P_1(k)Y_0(k)\P_2(k)
=\ma e^{2\g\Re k}O(1)&e^{\g(\Im k+\Re k)}O(1)\\
e^{\g(\Im k+\Re k)}O(1)&e^{2\g\Im k}O(1)\am .
$$
The asymptotics \er{asA1B1} gives
the asymptotics \er{asB1im}.
\BBox

\subsection{Asymptotics of resonances}
Now we determine the sharp asymptotics of the scattering
matrix $S(k)$ and the function $\det\O(k)$ in $\K_1$.

\begin{lemma}
Let $(p,q)\in\cH_1\ts\cH_0$ and let $p_+=p(\g-0)\ne 0$. Then
\[
\lb{asS1}
S(k)=1+O(k^{-1})-{p_+\/4k^2}e^{-i2k\g}\big(1+o(1)\big),
\]
as $k\in\ol\K_1,|k|\to\iy$, uniformly in $\arg k\in[0,{\pi\/2}]$,
\[
\lb{asD1D}
\det\O(k)={p_+\/4k^2}e^{2k\g}\big(1+o(1)\big)
+\Big({p_+\/4k^2}\Big)^2e^{2(1-i)k\g}\big(1+o(1)\big)
\]
as $k\in\ol\K_1^+,|k|\to\iy$,
uniformly in $\arg k\in[0,{\pi\/4}]$.
\end{lemma}

\no {\bf Proof.}
Let $k\in\ol\K_1,|k|\to\iy$.
Substituting the asymptotics \er{asA1im}
into the definitions
\er{S}--\er{ScA}
we obtain
\[
\lb{asS101}
S(k)=1+c_k\cA_0+{e^{2\g\Im k}O(1)\/k^3},\qqq
c_k={\pi\/i2k^3}.
\]
Substituting the asymptotics \er{ascA0imp} into \er{asS101}
we obtain the asymptotics \er{asS1}.

The definition \er{dcB1} gives
\[
\lb{dX}
\det\O=1+c_k\Tr(\O_0-\O_1)+c_k^2\det(\O_0-\O_1).
\]
Let $k\in\ol\K_1^+,|k|\to\iy$.
The identity \er{idcB01} and the asymptotics
\er{asB1im} give
$$
\Tr(\O_0(k)-\O_1(k))=i\cA_0(ik)+\cA_0(k)+e^{2\g\Re k}O(1).
$$
The asymptotics \er{ascA0imp} and \er{ascA0impi} imply
\[
\lb{asTrOm}
\Tr(\O_0(k)-\O_1(k))
={ip_+k\/2\pi}e^{2k\g}\Big(1+o(1)+e^{2\g(\Im k-\Re k)}O(1)\Big),
\]
Moreover, the identity \er{idcB01} and the asymptotics
\er{asB1im} yield
\[
\lb{dX1}
\begin{aligned}
\det(\O_0(k)-\O_1(k))=
\big(i\cA_0(ik)+e^{2\g\Re k}O(1)\big)
\big(\cA_0(k)+e^{2\g\Im k}O(1)\big)
\\
-i\big(\mB(k)+e^{\g(\Im k+\Re k)}O(1)\big)^2.
\end{aligned}
\]
The asymptotics \er{ascA0imp}, \er{ascA0impi} and \er{asmBimp}
give
$$
\begin{aligned}
i\cA_0(ik)+e^{2\g\Re k}O(1)
={ip_+k\/2\pi}e^{2k\g}\big(1+o(1)\big),\\
\cA_0(k)+e^{2\g\Im k}O(1)=
-{ip_+k\/2\pi}e^{-i 2k\g}\big(1+o(1)+e^{-2\g\Im k}O(k)\big),\\
\mB(k)+e^{\g(\Im k+\Re k)}O(1)=
-{ip_+k\/2\pi}(1+i)e^{(1-i) k\g}
\big(1+o(1)+e^{-2 \g\Im k}O(1)\big).
\end{aligned}
$$
Substituting these asymptotics into the relation \er{dX1} we obtain
\[
\lb{asDetOm}
\det(\O_0(k)-\O_1(k))
=-\Big({p_+k\/2\pi}\Big)^2e^{2(1-i)k\g}
\big(1+o(1)+e^{-2 \g\Im k}O(k)\big).
\]
Substituting the asymptotics \er{asTrOm}
and \er{asDetOm} into the identity \er{dX}
we obtain
$$
\begin{aligned}
\det\O(k)=1+{p_+e^{2k\g}\/4k^2}
\big(1+o(1)+e^{2\g(\Im k-\Re k)}O(1)\big)
\\
+\Big({p_+\/4k^2}\Big)^2e^{2(1-i)k\g}
\big(1+o(1)+e^{-2 \g\Im k}O(k)\big)
\\
=\Big({p_+\/4k^2}\Big)^2e^{2(1-i)k\g}
\Big(1+o(1)+{4k^2\/p_+}e^{i2k\g}\big(1+o(1)\big)\Big),
\end{aligned}
$$
which yields the asymptotics \er{asD1D}.
\BBox

\medskip

We are ready to determine asymptotics of resonances.

\medskip

\no {\bf Proof of Theorem \ref{ThAsCF}.}
The function $D(k)$ has a finite number of zeros in the domain
$\K_1$.

Let $k\in\K_1,|k|\to\iy$ and let $ik$ be a resonance.
The identity \er{cS(k)} shows that
$k$ is a zero of the function $S(k)$ in $\K_1$.
The asymptotics \er{asS1} and
the identity
$S(k)=0$ imply that $k$ satisfies
the equation
$$
k^2e^{i2k\g}={p_+\/4}
\big(1+o(1)\big).
$$
Then $k$ lies on the logarithmic curve $\G$ in $\K_1$, given by
$$
|k|={|p_+|^{1\/2}e^{\g\Im k}\/2}\big(1+o(1)\big),
$$
and satisfies
\[
\lb{asriK+}
ik={ij_n\pi\/\g}-{\log k\/\g}+{1\/2\g}\log {|p_+|\/4}+o(1)
\]
and there are no any other large resonances in $i\K_+$.

Let $k\in \K_1^+$, let $-k$ be a resonance and let $|k|$ be large
enough.
The identity \er{cT(k)} shows that $-k$
is a zero of the function $\det\O(k)$ in $\K_1$.
The identity $\det\O(k)=0$ and the asymptotics \er{asD1D} show that
$$
k^2e^{2ik\g}
=-{p_+\/4}\big(1+o(1)\big).
$$
Then $k$ lies on the curve $\G$
and satisfies
\[
\lb{asriK-}
-k=-{(j_{n}+{1\/2})\pi\/\g}-{i\log k\/\g}
+{i\/2\g}\log {|p_+|\/4}+o(1)
\]
and there are no any other large resonances in $-\K_1^+$.
The asymptotics \er{asriK+}, \er{asriK-} give
\er{asresiK+}, which yields
the asymptotics \er{asNj}.
\BBox

\subsection{Further discussions}
\lb{SectFD}

An entire function $f(z)$ is said to be of exponential type if there
is a constant $A$ such that $|f(z)|\leq\const e^{A|z|}$ everywhere.
The infimum of the set of $A$ for which such inequality   holds is
called the type of $f$. For each exponential type function $f$ we
define the types $\r _{\pm}(f)$ in $\C_\pm $ by
$$
\r _{\pm}(f)\equiv \lim \sup_{y\to \iy} {\log |f(\pm iy)|\/y} .
$$
 We introduce the class of exponential type functions. The function
$f$ is said to  belong    to the Cartwright class $\cC_\r$ if $f$ is
entire, of exponential type, and the following conditions  hold
true:
$$
\int_\R\frac{\log(1+|f(x)|)}{1+x^2}dx
<\infty,\qq\rho_+(f)=0,\qq\rho_-(f)=2\rho>0,
$$
for some $\rho>0$.
 We recall the Levinson Theorem  (see \cite{Ko88}):
{\it  Let the entire function $f\in \cC_\r$. Let $\cN(r)$ be the
number of zeros of the function $f$ in the disc $|k|<r$, counted
with multiplicity. Then $\cN (r,f)={2\r\/ \pi }r+o(r)$ as $ r\to \iy
$.}

We will discuss what properties of resonances of the second and
fourth order with compactly supported coefficients are common and
which are specific.

$\bu $ Common properties:

1) The determinants $D(k)$ and $d(k)$ are exponentially type
functions in terms of the variable $k$ (not $\l$) and each of them
has an axis of symmetry.

2) The resonances have the logarithmic type asymptotics for
coefficients with steps.

However, there are significant differences between the resonances
for Scr\"odinger operators and for fourth order operators.
Indeed, it is well known that the resonances for the Scr\"odinger
operators satisfy:

 $\bu $ specific  properties of $d(k)$

1) The Riemann surface for the determinant $d(\l^{1\/2})$ is the
Riemann surface for the function $\l^{1\/2}$. Then it has (as the
function of $\l$) two sheets and  $d(k)\sim 1$ as $|\l|\to \iy$ on
the first (physical) sheet and $d(k)\sim e^{2\g|\Im \sqrt \l|}$ on
the second (non-physical) sheet. There is a finite number of
eigenvalues on the first sheet and an infinite number of resonances
on the second one.

2) The determinant $d(k)$ belongs to the Cartwright class $\cC_\g$.
Then the Levinson Theorem describes the distribution of resonances
in the large disc.

3) The number of resonances $\cN_r$ in the disk $|\l|^{1\/2}<r$ has
asymptotics $\cN_r={2\g \/\pi}r(1+o(1))$ as  $r\to \iy$.

4) In order to obtain an analytic extension of the determinant from
the first sheet onto the second one uses one identity \er{cS(k)2o}.

$\bu $ specific  properties of the determinant $D$  for the fourth
order operator:

1)  The Riemann surface for the determinant $D(\l^{1\/4})$ is the
Riemann surface for the function $\l^{1\/4}$. Thus one has four
sheets:  $D\sim 1$ at $|\l|\to \iy$ on the first sheet, $D\sim
e^{2\g|\l|^{1\/4}}$ on the second and fourth sheets and $D\sim
e^{2\sqrt2\g|\l|^{1\/4}}$ on the third sheet. There is a finite
number of eigenvalues on the first (physical) sheet and an infinite
number of resonances on the other (non-physical) sheets. The number
of resonances in the large disc on the third sheet is, roughly
speaking, in two times more than on the second (or fourth) sheet.

2) The determinant $D$ is not in the Cartwright class.

3) The number of resonances $\cN_r$ in the disk $|\l|^{1\/4}<r$ has
asymptotics $\cN_r={4\g\/\pi}r(1+o(1))$ as  $r\to \iy$.

4) In order to obtain an  analytic extension of the determinant from
the first sheet onto the other sheets we need to use two identities
\er{cS(k)}, \er{cT(k)}.

\medskip

\no {\bf Acknowledgments.} \footnotesize  The study was
supported by the RFBR grant  No 16-01-00087.

\end{document}